\documentclass[reprint, aps, prl, footinbib]{revtex4-1}
\usepackage{amssymb,amsmath}
\usepackage{graphicx}
\usepackage{dcolumn}
\usepackage{bm}
\usepackage{braket}
\usepackage[obeybold]{siunitx}

\renewcommand{\Re}{\operatorname{Re}}
\renewcommand{\Im}{\operatorname{Im}}
\DeclareMathOperator{\Tr}{Tr}
\DeclareSIUnit\sample{Sa}

\begin{document}
\raggedbottom

\title{High-fidelity software-defined quantum logic on a superconducting qudit}
\author{Xian Wu}
\email{Correspondence to: wu47@llnl.gov}
\affiliation{Lawrence Livermore National Laboratory, Livermore, CA 94550, USA}

\author{S. L. Tomarken}
\affiliation{Lawrence Livermore National Laboratory, Livermore, CA 94550, USA}

\author{N. Anders Petersson}
\affiliation{Lawrence Livermore National Laboratory, Livermore, CA 94550, USA}

\author{L. A. Martinez}
\affiliation{Lawrence Livermore National Laboratory, Livermore, CA 94550, USA}

\author{Yaniv J. Rosen}
\affiliation{Lawrence Livermore National Laboratory, Livermore, CA 94550, USA}

\author{Jonathan L DuBois}
\affiliation{Lawrence Livermore National Laboratory, Livermore, CA 94550, USA}
\date{\today}

\begin{abstract}
We present an efficient approach to achieving arbitrary, high-fidelity control of a multi-level quantum system using optimal control techniques. As an demonstration, we implement a continuous, software-defined microwave pulse to realize a $0 \leftrightarrow 2$ SWAP gate that achieves an average gate fidelity of $\SI{99.4}{\percent}$. We describe our procedure for extracting the system Hamiltonian, calibrating the quantum and classical hardware chain, and evaluating the gate fidelity. Our work represents an alternative, fully generalizable route towards achieving universal quantum control by leveraging optimal control techniques.
\end{abstract}

\maketitle

Due to recent breakthroughs in quantum technology \cite{chow:2013, Chen:2014, Kelly:2015aa, Arute:2019aa, Reagor:2018,Debnath:2016aa,Wright:2019aa}, quantum information is entering the Noisy Intermediate-Scale Quantum (NISQ) era \cite{Preskill:2018}. In the NISQ era, quantum processors consisting of $50$-$100$ qubits and lacking fault-tolerant noise correction protocols have the potential to perform tasks surpassing today's best classical computers \cite{Bernien:2017aa,Nam:2019aa,Arute:2019aa}. Most recent efforts to reduce computational error rates have focused on extending qubit coherence times by leveraging improvements in fabrication and qubit design, e.g.~Refs.~\onlinecite{Nersisyan:2019, Nguyen:2019, Alex:2020}. However, a less explored route lies in harnessing classical computing power to co-design efficient quantum control protocols to extend the computational limits of state-of-the-art quantum processors \cite{Khaneja:2005,Floether:2012,Holland:2019}. One promising avenue is quantum optimal control \cite{Khaneja:2005,Machnes:2011,Cheng:2020},  where nonlinear optimization techniques are utilized to design microwave pulses that can perform arbitrary unitary operations. This allows for a departure from traditional discrete gate-set approaches where quantum states are controlled using a reduced set of `primitive' single- and two-qubit gates.

In the NISQ era, one of the most anticipated applications of quantum computation is quantum simulation \cite{Paraoanu}, which often requires continuous control over complex, multi-qubit time-evolution.  Currently, reduced gate-set fidelities $\SI{\ge99}{\percent}$ are routinely achieved for single-qubit gates \cite{Barends:2014aa,Sheldon:2016,Rol:2017,Reagor:2018} and, more recently, for several two-qubit gates \cite{Dewes:2012, Chen:2014, Sheldon:2016, Rosenblum:2018aa, Hong:2020,Kjaergaard:2020b}. However, fixed gate-set approaches tend to perform poorly for quantum simulation because complex quantum system evolution can only be achieved by concatenating large numbers of primitive gates, resulting in very deep quantum circuits with significant cumulative error rates \cite{Lanyon57, Barends:2015, OMalley:2016, Kandala:2017aa}. In contrast with reduced gate-set approaches, employing quantum optimal control methods allows one to construct a single custom gate that realizes a target evolution directly. Complex gates constructed with the aid of optimal control algorithms often require net shorter durations and yield higher process fidelities than is possible with composite gate sequences \cite{Holland:2019, shi2020quantum}. Before one can implement an optimal control gate, the device and control Hamiltonian must be understood in detail for optimizing control function that realize target operation. This requires a detailed characterization of both the physical quantum hardware and the quantum-classical transfer function arising from room temperature electronics, cryogenic control lines, and related components. Once the device Hamiltonian and transfer function are known, no gate-level hardware calibration is required for the application of arbitrary optimal control gates, making this approach highly flexible.

In this work, we report our experimental effort towards implementing quantum optimal control on a superconducting transmon qudit to realize a nontrivial unitary operator. By extending the dimension of the Hilbert space to a $d$-level \emph{qudit}, we achieve a computational advantage with fewer physical devices \cite{Neeley:2009,Lanyon:2009aa,chow:2013,Helle:2000,Cerf:2002}. Previous studies on multi-level transmon qudits indicate that the higher energy levels are a useful quantum resource \cite{Bianchetti:2010, Peterer:2015, Rosenblum266, Blok:2020}. In this manuscript, we present full characterization of the lowest four energy levels of a three-dimensional (3D) transmon device \cite{Paik:2011} and the quantum-classical transfer function that underlies the calculation of the optimal control gate. We perform both single-gate and repeated-gate measurements. We demonstrate an efficient characterization of the process matrix and estimate an average gate fidelity of $\SI{99.4}{\percent}$.

The system's Hamiltonian in the presence of time-dependent control drives is given by $H(t) = H_0 + \sum_{j=1}^N u_j(t) H_j$ where $H_0$ describes the time-independent Hamiltonian and the $H_j$ are the available control Hamiltonians. The $u_j(t)$ is time varying amplitudes for each control term. Using numerical optimization \cite{Khaneja:2005, Lloyd:2014}, we can find a set of $u_j(t)$ that realize the target unitary transformation $U_{\text{targ}}$, within an acceptable error, according to
\begin{equation}
	U_{\text{targ}} = \mathcal{T} e^{-i\int_{0}^{T_g}H(t)dt/\hbar},
\end{equation}
where $\mathcal{T}$ represents time-ordering of the exponential, $T_g$ is the gate time, and $\hbar$ is Planck's reduced constant.

\begin{figure}[b]
	\includegraphics[width = 0.48\textwidth]{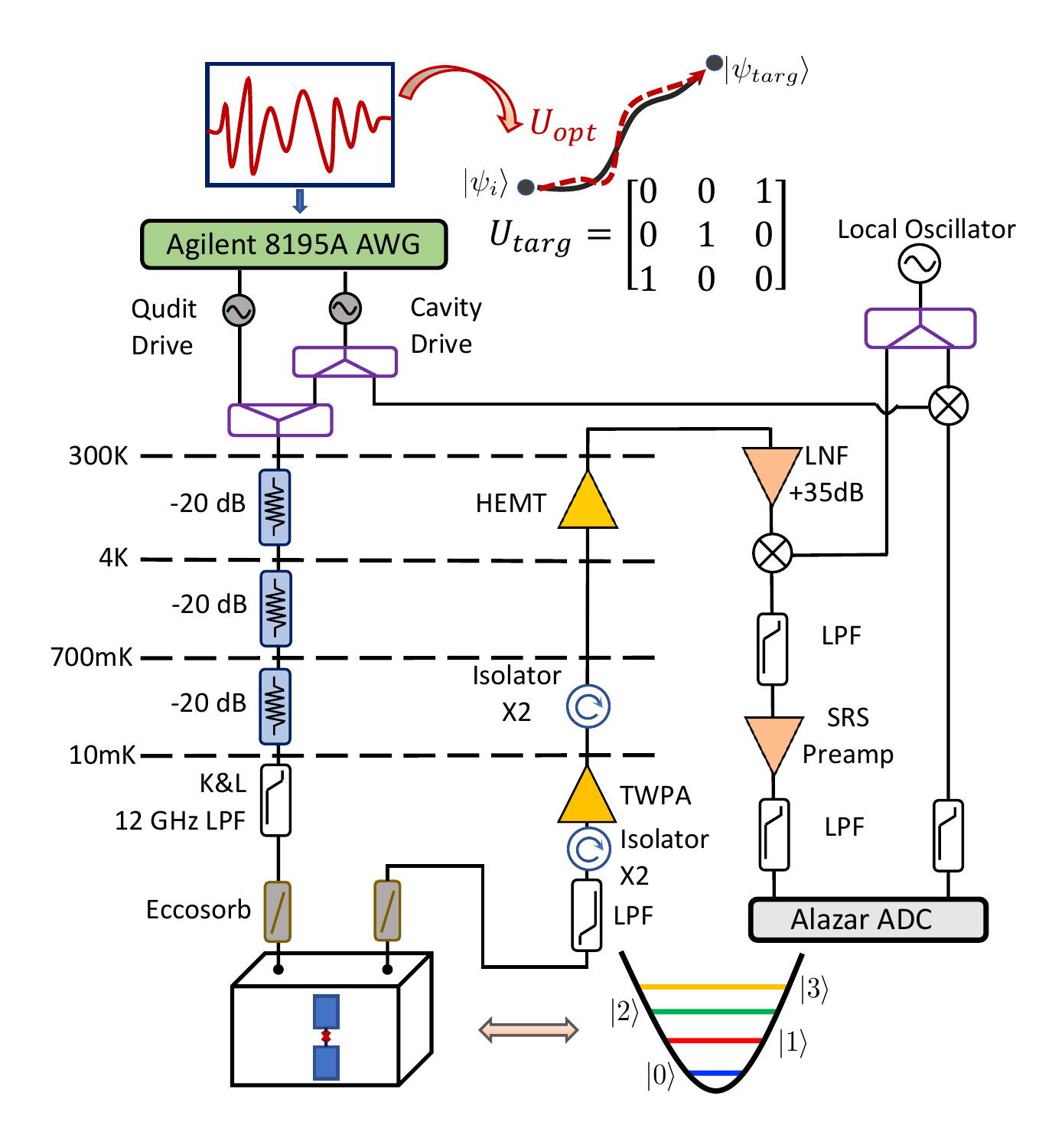}
	\caption{\textbf{Experiment setup for implementing an optimized control pulse on a transmon qudit.} The top cartoon illustrates the optimization procedure for a qudit control pulse given a target unitary operator. The lower schematic depicts the measurement diagram.}
	\label{fig:1}
\end{figure}

For the purposes of demonstrating the optimal control technique, we have chosen a $0\leftrightarrow2$ SWAP gate as the $U_{\text{targ}}$, which has applications in typical computations and experiments and is more complex to construct than a standard single-qubit operation. For qutrits or qudits, this SWAP gate can be used to prepare initial states or perform fast reset on states after measurement. The first three levels $\ket{0}, \ket{1}, \ket{2}$ of our transmon qudit make up the computational space while the fourth level $\ket{3}$ monitors the leakage out of the computational space. The transmon is coupled to a 3D aluminum superconducting cavity which is used to dispersively read out the qudit state \cite{Blais:2004}. The experimental setup is shown in Fig.~\ref{fig:1}. Parameters of the transmon  and cavity are listed in Table~\ref{tab:1}. 

\begingroup
\squeezetable
\begin{table}[b]
\caption{\label{tab:1}Device parameters}
\begin{tabular}{l|l}
\rule{0pt}{3ex}\textbf{Parameter} & \textbf{Value}\\
\colrule	
\rule{0pt}{3ex} $\ket{0} - \ket{1}$ transition frequency $\omega_q^{(0,1)}/2\pi$ & $\SI[parse-numbers=false]{4.09948}{\giga\hertz}$\\
Relaxation time $T_1$ for the $\ket{0} - \ket{1}$ transition  & $\SI{55}{\micro\second}$\\
Ramsey\footnotemark[1]\footnotetext[1]{\tiny Ramsey measurements for the $1$-$2$ and $2$-$3$ transitions show aliasing similar to Ref.~\onlinecite{Peterer:2015}, making it difficult to accurately extract $T_2^*$. Instead, $T_2^E$ is reported. Decay times estimated from Ramsey data before aliasing behavior occurs are used in the Lindbladian terms in simulations. ($1$-$2$: $\SI{3.838}{\micro\second}$, $2$-$3$: $\SI{0.224}{\micro\second}$)} decay time $T_2^*$ for the $\ket{0} - \ket{1}$ transition  & $\SI{35}{\micro\second}$\\
$\ket{1} - \ket{2}$ transition frequency $\omega_q^{(1,2)}/2\pi$ & $\SI[parse-numbers=false]{3.87409}{\giga\hertz}$\\
Relaxation time $T_1$ for the $\ket{1} - \ket{2}$ transition  & $\SI{26}{\micro\second}$\\
Echo\footnotemark[1] time $T_2^E$ for the $\ket{1} - \ket{2}$ transition  &  $\SI{13}{\micro\second}$\\
$\ket{2} - \ket{3}$ transition frequency $\omega_q^{(2,3)}/2\pi$ & $\SI[parse-numbers=false]{3.61938}{\giga\hertz}$\\
Relaxation time $T_1$ for the $\ket{2} - \ket{3}$ transition  & $\SI{18}{\micro\second}$\\
Echo time\footnotemark[1] $T_2^E$ for the $\ket{2} - \ket{3}$ transition  &   $\SI{7.5}{\micro\second}$\\
Readout resonator frequency $\omega_r/2\pi$  & $\SI[parse-numbers=false]{7.0768}{\giga\hertz}$ \\
Effective dispersive coupling strength, $\chi_{qc}/2\pi$ & $\SI[parse-numbers=false]{1.017}{\mega\hertz}$\\
\colrule
\end{tabular}
\end{table}
\endgroup

The transmon's Hamiltonian is given by $H_0 = \hbar\sum_{k=0}^3\omega_k \ket{k} \bra{k}$ where $\hbar\omega_{k}$ and $\ket{k}$ are the eigenenergy and eigenstate of the transmon's $k$th level. We set $\hbar = 1$ for further equations in this manuscript. Because the optimized control pulse must drive qudit transitions at roughly $\SI{4}{\giga\hertz}$, the dominant frequency components of the control pulse will be about $\SI{3}{\giga\hertz}$ or more detuned from the readout cavity. As a result, in both the drive Hamiltonian and interaction Hamiltonian (below), we can safely neglect terms associated with the readout cavity. 

Next, we derive the control Hamiltonian $H_j$ for the qudit being driven by a microwave pulse. The time-dependent Hamiltonian that describes the interaction between the electric field inside the cavity and a multi-level quantum system is \cite{Gerry}
\begin{align}
	\begin{split}
		H_{\text{int}} &= \sum_{j=1}^N u_jH_j = (c + c^{\dagger})(\xi e^{-i\omega_d t}+\xi^*e^{i\omega_d t})\\        
		&=(c + c^{\dagger})(2\Re(\xi)\cos(\omega_d t)+2\Im(\xi)\sin(\omega_d t))
	\end{split}
\end{align}
where $c$ $\left(c^{\dagger}\right)$ is the lowering (raising) operator in the transmon eigenbasis, $\xi$ is the drive strength, and $\omega_d$ is the drive frequency.In order to extract the raising and lowering operators, we model the transmon as a Cooper pair box in the charge basis and explicitly calculate the raising and lowering operators for a truncated charge basis \footnote{See the supplemental material at [PRL LINK] which contains Refs.~\onlinecite{Blais:2004,scikit-learn, Reagor:2018,Gerry} for details of the model and calculation.}.
    
To slow down the time scales in the numerical optimization, we transform into the rotating frame of the drive using the unitary operator $R(t)~=~\exp\left(\left[i\omega_d t\left( \sum_k k \ket{k}\bra{k} \right) \right]\right)$ yielding
\begin{align}
	\begin{split}
		H_{\text{rot}} &= R(t)\left(H_0 + H_{\text{int}}\right)R(t)^{\dagger} - iR(t)\dot{R}(t)^{\dagger} \\
            &= \sum_k \Delta_{k} \ket{k} \bra{k} + \Re(\xi)\left(\tilde{c} + \tilde{c}^{\dagger}\right) - i\Im(\xi)\left(\tilde{c} - \tilde{c}^{\dagger}\right)
	\end{split}
	\label{labopt}
\end{align}
where $\Delta_k \equiv \omega_k - k \omega_d$, and $\tilde{c}$ $\left(\tilde{c}^{\dagger}\right)$ is the lowering (raising) operator in the rotating frame.Without loss of generality, the ground state energy is defined to be zero, $\omega_0 = 0$. We choose $\omega_d = \omega_1$, and make use of the rotating wave approximation wherein terms oscillating at $\pm2\omega_d$ are neglected. In the rotating frame we have two $H_j$ and two $u_j$ to be optimized:
\begin{gather}
	\begin{aligned}
		H_{1} &= \tilde{c} + \tilde{c}^{\dagger}
		& H_{2} &= -i\left(\tilde{c} - \tilde{c}^{\dagger}\right)
		\\
		u_{1} &= \Re(\xi)
		& u_{2} &= \Im(\xi).\label{rwaopt}
	\end{aligned}
\end{gather}
As shown in Eq.~\ref{rwaopt}, the time-dependence on the control functions $u_1$, $u_2$ has been relieved. Because we have chosen to optimize the control pulse in the rotating frame, the target operator becomes $R\left(T_g\right)U_{\text{targ}}R^{\dagger}(0)$.

Now, we turn to the numerical optimization. There are multiple software packages available for finding optimal control pulses \cite{Johansson:2013, Leung:2017}. Most implementations rely on gradient-descent methods \cite{Khaneja:2005, Fouquieres:2011} where the continuous gate can be subjected to experimentally relevant constraints such as maximal drive amplitude and gradient as well as boundary conditions of the control pulse. Here, we use the recently developed package described in Ref.~\onlinecite{anders:2020} because it gives precise control over the dominant frequencies in the spectra of the control functions. The optimizer minimizes the objective function ${\cal G}$ defined as below:
\begin{gather}
\begin{aligned}
	{\cal G} &= \left(1 - F_g^2\right) + \frac{1}{T_g}\int_0^{T_g} \mbox{Tr}\left(U_{\text{opt}}^\dagger(t) W U_{\text{opt}}(t)\right)\, dt\\
	F_g &= \frac{1}{d} \left| \mbox{Tr}\left(U_{\text{targ}}^\dagger U_{\text{opt}}\left(T_g\right)\right) \right|,~~
	W = \left(\begin{smallmatrix}
  		0 &&&\\
  		& 0 &&\\
 		&& 0 &\\
		&&& 1 \end{smallmatrix}\right)
\end{aligned}\label{obj}
\end{gather}
where $U_{\text{targ}}$ is the target unitary operator in the rotating frame. $U_{\text{opt}}(t)$ is the propagator based on the time-dependent control solution and is used to calculate the gate fidelity $F_g$ for the generated control solution. The second term in Eq.~\ref{obj} calculates the $\ket{3}$ population throughout the gate which is used to minimize the leakage outside the computational space.

$T_g$ is chosen to be $\SI{150}{\nano\second}$ to comply with the maximum drive amplitude we can apply in our measurement setup. Figure~\ref{fig:1} contains the measurement circuit details. A Keysight 8195A arbitrary waveform generator (AWG) directly synthesizes the control drive with an output limit of \SI{1}{\volt}. Taking into account the total attenuation of the input line, the drive strength is limited to about $\SI{6}{\mega\hertz}$ in the rotating frame (or $\SI{12}{\mega\hertz}$ in the laboratory frame). $T_g$ may be further reduced by decreasing the line attenuation, increasing the AWG output limit, or reducing the coupling quality factor for signals entering the qudit cavity. See the SM for details. We set the time resolution of the control pulse to $\SI{1/32}{\nano\second}$ to match the $\SI{32}{\giga\sample\per\second}$ sampling rate of the AWG. Figure S2(b) in the SM shows the optimized control functions in both the rotating and laboratory frames with calculated $F_g=\SI{99.997}{\percent}$.

\begin{figure}
	\includegraphics[width = 0.48\textwidth]{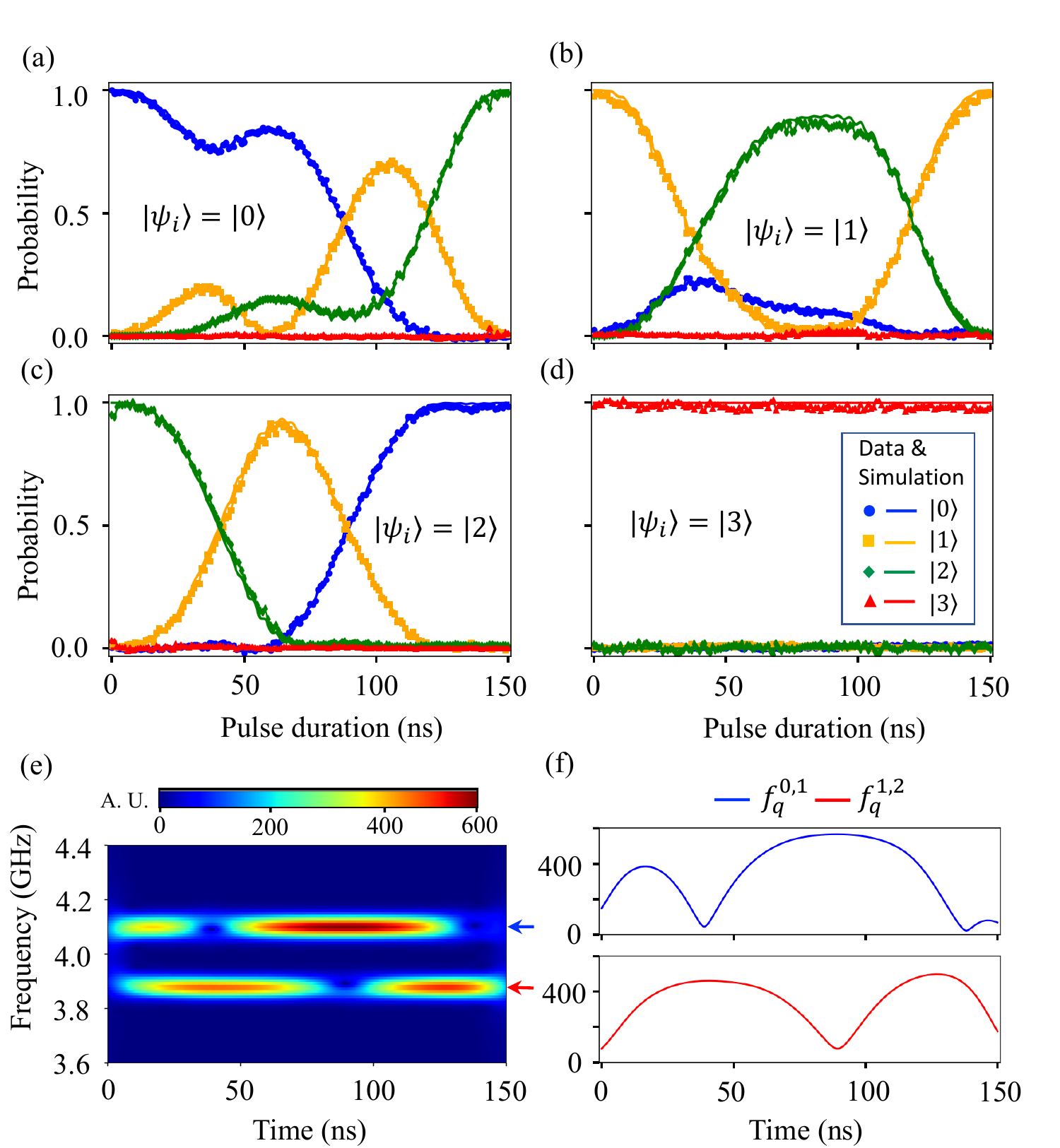}
	\caption{\textbf{Measuring the qudit state probabilities during a single gate application and time-frequency analysis of the control pulse}
	\textbf{(a)}--\textbf{(d)} Initial states are prepared in $\ket{0}, \ket{1}, \ket{2}, \ket{3}$ for each gate application. 
	 $\ket{1}$ is prepared with a square pulse at $\omega_q^{(0,1)}$.  $\ket{2}$ and $\ket{3}$ are prepared with a sequence of square pulse at $\omega_q^{(0,1)}, \omega_q^{(1,2)}$ and $\omega_q^{(2,3)}$. Solid lines are master equation simulations for ideal gate implementations. \textbf{(e)} Wavelet transformation of the laboratory frame control pulse. \textbf{(f)} Linecuts of (e) at $f_q^{(0,1)}$ and $f_q^{(1,2)}$.}
	\label{fig:3}
\end{figure}

In order to investigate the $0\leftrightarrow2$ SWAP gate, we perform both single- and repeated-gate measurements. First, we characterize the behavior of a single gate by measuring the state occupation probabilities for the lowest four qudit states throughout the duration of the gate. We initialize the qudit state separately to $\ket{0},\ket{1},\ket{2},\ket{3}$ and measure the occupation probability for each state at $\SI{1}{\nano\second}$ intervals as shown in Fig.~\ref{fig:3}(a)--(d). We perform single-shot readout at the end of each measurement and infer the final state of the qudit through classification (see SM for details). In order to evaluate the performance of a single gate, we simulate the qudit dynamics using the master equation formalism with the QuTiP python library \cite{Johansson:2013}. Because the total gate time is much shorter than the coherence times of all of the states involved, Lindbladian terms \cite{Lindblad} associated with relaxation and dephasing are not included in the simulations shown as solid lines in Fig.~\ref{fig:3}. We observe good agreement between our measurements and simulations.

The qudit probability evolution tracks the dominant frequency modes of the control signal throughout the duration of the pulse and can be used to study the dynamics of the control pulse which often contains a broad mixture of frequency components and is difficult to intuit in the time-domain directly. Whenever occupation between adjacent states $\ket{i}$ and $\ket{i+1}$ is exchanged, the frequency component at $f_q^{(i,i+1)} = \omega_q^{(i,i+1)}/2\pi$ of the drive has significant amplitude. For example, population exchange between states $\ket{0}$ and $\ket{1}$ at the beginning of the control pulse in Fig.~\ref{fig:3}(a) and (b) indicates a strong spectral contribution from $f_q^{(0,1)} = \omega_q^{(0,1)}/2\pi$ until about $\SI{40}{\nano\second}$ where a small plateau of $\ket{0}$ develops. Later, the $f_q^{(0,1)}$ drive is resumed with stronger intensity until about $\SI{120}{\nano\second}$, after which the $\ket{0}$ population becomes essentially flat. A similar analysis applies to the $f_q^{(1,2)}$ Fourier component and the $\ket{1}$ and $\ket{2}$ state occupation in Fig.~\ref{fig:3}(b) and (c). Importantly, Fig.~\ref{fig:3}(d) verifies that there is no drive component near frequency $f_q^{(2,3)}$. Our qudit analysis of the control pulse is consistent with time-frequency analysis of the control pulse. Morlet wavelet analysis with $150$ cycles \cite{Morlet} of the control signal is presented in Fig.~\ref{fig:3}(e) and (f) and verifies that the dominant frequency components are centered around  $f_q^{(0,1)}$ and $ f_q^{(1,2)}$ and that their amplitudes vary in time in agreement with the qudit population trajectories in Fig.~\ref{fig:3}(a)--(d). In Fig.~\ref{fig:3}(e), the population of the qudit is moved in two major steps throughout the pulse. The drive is attempting to use the $\ket{1}$ state as an intermediary between the $\ket{0}$ and $\ket{2}$ state populations, while restoring the $\ket{1}$ state at the end of the drive. 
Previous work has demonstrated an adiabatic process (STIRAP) that achieves coherent transfer of population from the $\ket{0}$ state to the $\ket{2}$ state without populating the $\ket{1}$ state. \cite{Kumar:2016aa, Xu:2016aa}. 
With optimal control, a similar control solution can be found, which typically requires longer gate time.

\begin{figure}[ht!]
	\includegraphics[width = 0.48\textwidth]{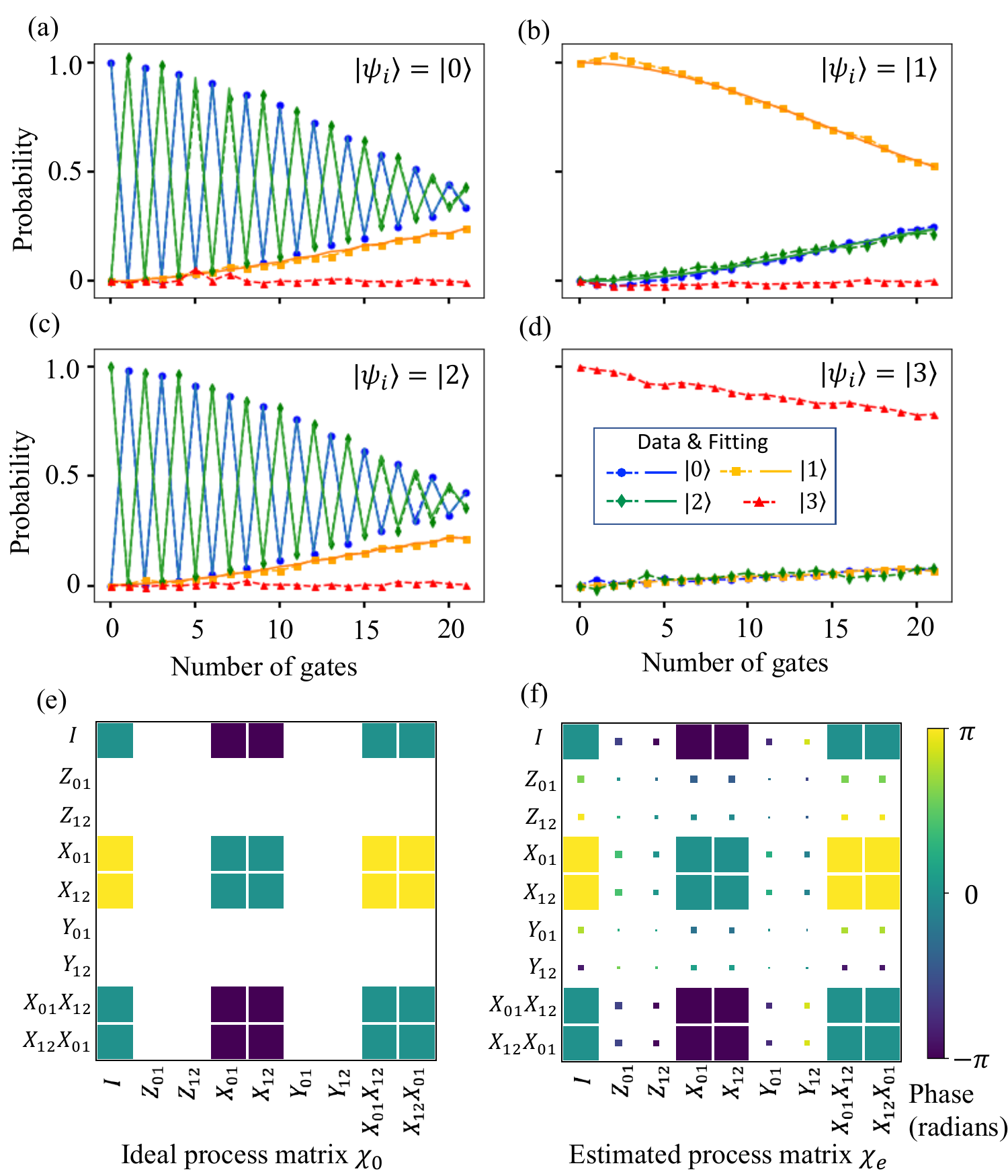}
	\caption{
	\textbf{Measurement of state evolution during repeated gate application and process matrix analysis} \textbf{(a)}--\textbf{(d)} Measured state probabilities for multiple gate applications when preparing the initial state in $\ket{0}, \ket{1}, \ket{2}, \ket{3}$. Symbols connected with dashed lines are measurement results. Solid lines are simulations of the expected measurement outcome with the estimated process matrix.  \textbf{(e),(f)} Hinton diagrams of the ideal process matrix and estimated process matrix for the implemented $0\leftrightarrow2$ SWAP gate. The magnitude and phase for each matrix element are represented by its area and color, respectively.}
	\label{fig:4}
\end{figure}

Finally, we analyze the gate performance by measuring the qudit state occupation upon repeated application of the optimal control gate up to $21$ times. Measurements are performed for all four initial states with results presented in Fig.~\ref{fig:4}(a)--(d). For initial states $\ket{0}$ and $\ket{2}$, coherent state population exchange between $\ket{0}$ and $\ket{2}$ is observed. There is significant leakage into $\ket{1}$, while leakage into $\ket{3}$ is minimal.  For the initial states $\ket{1}$ and $\ket{3}$, no obvious coherent behavior is observed, as expected for a $0\leftrightarrow2$ SWAP gate. As the number of gate applications increases, the amplitude of the $\ket{0}$ and $\ket{2}$ population transfer decreases, an effect that we attribute to both relaxation and decoherence intrinsic to the physical device and coherent control errors (imperfect implementation of the control pulse). Additionally, there are errors associated with state preparation and state readout, however, we believe these latter contributions are much less significant due to the successful single-gate measurements presented in Fig.~\ref{fig:3}.

To examine the device's intrinsic error rate, we simulate the qudit response under repeated applications of the control pulse with the measured decay and decoherence rates. The simulation result is plotted in Fig.~S3(a)--(d) of the SM and qualitatively agrees with our measurement results. The simulation for the $\ket{3}$ state is in good agreement, suggesting our estimation of the device's decay error rate ($T_1$) is accurate. Comparing to the simulation results, we observe more attenuation of $\ket{0}$ and $\ket{2}$ state population transfer and zigzag behavior of the $\ket{1}$ state population, suggesting the coherent control error is significant. To quantify this coherent error, we perform process tomography analysis on both repeated gate measurement results and simulation results. 

We extract the process matrix $\chi$ to describe the drive's effect on an arbitrary input state $\rho$. Following Ref.~\onlinecite{QPT}, we can choose one complete gate set to describe any process for a $d$-dimensional system:
\begin{align}
	\mathcal{E}(\rho)= &\sum_{m,n=0}^{d^2-1}\chi_{mn}B_m\rho B_n^{\dagger}\label{choi1}\\
	\begin{split}
		B_n = &\left\{I, Z_{01},Z_{12},X_{01},X_{12},Y_{01}\right.\\
          		&~\left.Y_{12},X_{01}X_{12},X_{12}X_{01}\right\}.
	\end{split} \label{choi2}
\end{align}
Here, $\{B_n\}$ forms a complete gate set to represent any $d\times d$ matrix. $\chi$ is a positive superoperator which completely characterizes the process $\mathcal{E}$ using the basis operators  $\{B_n\}$. The established procedure \cite{Chow:2009} to estimate the process matrix $\chi$ is to prepare many different initial states, apply the mapping once, and fit for the most likely $\chi$ based on the measurement outcomes. Here, we perform the estimation of $\chi$ using a different method. We prepare three initial states $\ket{0},\ket{1},\ket{2}$ and apply the mapping sequentially to obtain a series of state occupation probabilities. The best-fit $\chi$ matrix is estimated by minimizing the difference between the generated state probabilities for a given implementation of $\chi$ and the measurement results. We are motivated to use this method because quantum simulation often requires repeated application of the same gate \cite{Barends:2015,Holland:2019}, from which the process fidelity can be estimated without additional experiments. The caveat of this approach is that there is no guarantee that the set of generated states is sufficient to constrain the fitting. 

We examine the measured state populations in Fig.~\ref{fig:4}(a)--(c). We observe that states spanning a variety of basis elements are generated. This indicates we have likely generated a sufficient number of states to constrain the fit results. We believe that all states remain highly coherent because the measurement outcome is produced mainly by the control error (as opposed to state decay or dephasing). The ideal and estimated $\mathrm{\chi}$ matrices are plotted in Fig.~\ref{fig:4}(e) and (f) for comparison. Following Ref.~\onlinecite{QPT}, we calculate entanglement and gate fidelity according to equations below:
\begin{align}
F_e(\rho,U,\mathcal{E})= &\sum_{mn}\chi_{e,mn}\Tr\left(U^{\dagger}B_m\rho\right)\Tr\left(\rho B_n^{\dagger}U\right)\label{ent_fid}\\
F_g(\ket{\psi},U,\mathcal{E}) =&\bra{\psi} U^{\dagger}\mathcal{E(\ket{\psi}\bra{\psi})}U\ket{\psi} \label{gate_fid}
\end{align}
where $\chi_e$ is the estimated process matrix, $\rho$ represents the density matrix form of the input state, and $\ket{\psi}$ represents a pure state. The averaged entanglement fidelity is $\SI{99.2}{\percent}$ and the averaged gate fidelity is $\SI{99.4}{\percent}$. 
We repeat the same analysis of the simulation results containing only the device intrinsic error and extract an average gate fidelity of $\SI{99.6}{\percent}$. Therefore, we estimate the coherent control error to be about $\SI{0.2}{\percent}$.

We have demonstrated a $0\leftrightarrow2$ SWAP gate with a software-defined optimal control pulse on the lowest four levels of a 3D transmon qudit. The fidelity of our optimal control gate is comparable to two-qubit gates utilizing a small set of discrete gates \cite{Barends:2014aa}. However, our implemented control waveform is broadband and representative of the typical spectral width needed to realize an arbitrary unitary gate. For more complex unitary evolution (e.g.~Refs.~\onlinecite{Barends:2015, shi2020quantum}), the discrete gate-set fidelities suffer from the compounding error associated with the concatenation of $20-30$ standard gates. The main advantage of our approach is that once the control system is calibrated, we achieve gate fidelities comparable to our $0\leftrightarrow2$ SWAP gate even for complex multi-level control \cite{shi2020quantum} without additional gate-level calibration. This proof of principle demonstration can be generalized to multi-qubit systems as well to achieve arbitrary unitary transformations, offering an alternate route to achieving high-fidelity multi-qubit quantum control. Additionally, these optimal control techniques can be used in concert with discrete gate-set approaches to optimize the pulse shape of the constituent gates \cite{PhysRevA.82.040305}. Our results provide a guidepost for further development of optimal control techniques for many quantum information applications, particularly in the NISQ era.

\begin{acknowledgments}
This work was performed under the auspices of the U.S. Department of Energy by Lawrence Livermore National Laboratory under Contract No. DE-AC52-07NA27344.  X. W., S. T., L. M., Y. J. and J. D. acknowledge the Department of Energy Office of Advanced Scientific Computing Research, Quantum Testbed Pathfinder Program under Award No. 2017-LLNL-SCW1631 for support of development of quantum hardware and control methods used in this work. N. A. P. gratefully acknowledge the Laboratory Directed Research and Development program (Grant No. LDRD20-ERD-028) for supporting the development of quantum optimal control methods and software.
We thank James Colless, Irfan Sidiqqi, the Quantum Nanoelectronics Laboratory, and ARO/LPS for providing the transmon sample used in this work. We gratefully acknowledge support from the National Nuclear Security Administration Advanced Simulation and Computing Beyond Moores Law program (Grant No. LLNL-ABS-795437) for development and support of the LLNL quantum device integration testbed facility and Lab Directed Research and Development (Grant No. LDRD19-DR-005) for support of system calibration and data analysis. LLNL-JRNL-810657
\end{acknowledgments}

\bibliography{wuEtAl_optimalControl_PRL}

\providecommand{\noopsort}[1]{}\providecommand{\singleletter}[1]{#1}%
\begin{thebibliography}{55}%
\makeatletter
\providecommand \@ifxundefined [1]{%
 \@ifx{#1\undefined}
}%
\providecommand \@ifnum [1]{%
 \ifnum #1\expandafter \@firstoftwo
 \else \expandafter \@secondoftwo
 \fi
}%
\providecommand \@ifx [1]{%
 \ifx #1\expandafter \@firstoftwo
 \else \expandafter \@secondoftwo
 \fi
}%
\providecommand \natexlab [1]{#1}%
\providecommand \enquote  [1]{``#1''}%
\providecommand \bibnamefont  [1]{#1}%
\providecommand \bibfnamefont [1]{#1}%
\providecommand \citenamefont [1]{#1}%
\providecommand \href@noop [0]{\@secondoftwo}%
\providecommand \href [0]{\begingroup \@sanitize@url \@href}%
\providecommand \@href[1]{\@@startlink{#1}\@@href}%
\providecommand \@@href[1]{\endgroup#1\@@endlink}%
\providecommand \@sanitize@url [0]{\catcode `\\12\catcode `\$12\catcode
  `\&12\catcode `\#12\catcode `\^12\catcode `\_12\catcode `\%12\relax}%
\providecommand \@@startlink[1]{}%
\providecommand \@@endlink[0]{}%
\providecommand \url  [0]{\begingroup\@sanitize@url \@url }%
\providecommand \@url [1]{\endgroup\@href {#1}{\urlprefix }}%
\providecommand \urlprefix  [0]{URL }%
\providecommand \Eprint [0]{\href }%
\providecommand \doibase [0]{http://dx.doi.org/}%
\providecommand \selectlanguage [0]{\@gobble}%
\providecommand \bibinfo  [0]{\@secondoftwo}%
\providecommand \bibfield  [0]{\@secondoftwo}%
\providecommand \translation [1]{[#1]}%
\providecommand \BibitemOpen [0]{}%
\providecommand \bibitemStop [0]{}%
\providecommand \bibitemNoStop [0]{.\EOS\space}%
\providecommand \EOS [0]{\spacefactor3000\relax}%
\providecommand \BibitemShut  [1]{\csname bibitem#1\endcsname}%
\let\auto@bib@innerbib\@empty
\bibitem [{\citenamefont {Chow}\ \emph {et~al.}(2013)\citenamefont {Chow},
  \citenamefont {Gambetta}, \citenamefont {Cross}, \citenamefont {Merkel},
  \citenamefont {Rigetti},\ and\ \citenamefont {Steffen}}]{chow:2013}%
  \BibitemOpen
  \bibfield  {author} {\bibinfo {author} {\bibfnamefont {J.~M.}\ \bibnamefont
  {Chow}}, \bibinfo {author} {\bibfnamefont {J.~M.}\ \bibnamefont {Gambetta}},
  \bibinfo {author} {\bibfnamefont {A.~W.}\ \bibnamefont {Cross}}, \bibinfo
  {author} {\bibfnamefont {S.~T.}\ \bibnamefont {Merkel}}, \bibinfo {author}
  {\bibfnamefont {C.}~\bibnamefont {Rigetti}}, \ and\ \bibinfo {author}
  {\bibfnamefont {M.}~\bibnamefont {Steffen}},\ }\href {\doibase
  10.1088/1367-2630/15/11/115012} {\bibfield  {journal} {\bibinfo  {journal}
  {New Journal of Physics}\ }\textbf {\bibinfo {volume} {15}},\ \bibinfo
  {pages} {115012} (\bibinfo {year} {2013})}\BibitemShut {NoStop}%
\bibitem [{\citenamefont {Chen}\ \emph {et~al.}(2014)\citenamefont {Chen},
  \citenamefont {Neill}, \citenamefont {Roushan}, \citenamefont {Leung},
  \citenamefont {Fang}, \citenamefont {Barends}, \citenamefont {Kelly},
  \citenamefont {Campbell}, \citenamefont {Chen}, \citenamefont {Chiaro},
  \citenamefont {Dunsworth}, \citenamefont {Jeffrey}, \citenamefont {Megrant},
  \citenamefont {Mutus}, \citenamefont {O'Malley}, \citenamefont {Quintana},
  \citenamefont {Sank}, \citenamefont {Vainsencher}, \citenamefont {Wenner},
  \citenamefont {White}, \citenamefont {Geller}, \citenamefont {Cleland},\ and\
  \citenamefont {Martinis}}]{Chen:2014}%
  \BibitemOpen
  \bibfield  {author} {\bibinfo {author} {\bibfnamefont {Y.}~\bibnamefont
  {Chen}}, \bibinfo {author} {\bibfnamefont {C.}~\bibnamefont {Neill}},
  \bibinfo {author} {\bibfnamefont {P.}~\bibnamefont {Roushan}}, \bibinfo
  {author} {\bibfnamefont {N.}~\bibnamefont {Leung}}, \bibinfo {author}
  {\bibfnamefont {M.}~\bibnamefont {Fang}}, \bibinfo {author} {\bibfnamefont
  {R.}~\bibnamefont {Barends}}, \bibinfo {author} {\bibfnamefont
  {J.}~\bibnamefont {Kelly}}, \bibinfo {author} {\bibfnamefont
  {B.}~\bibnamefont {Campbell}}, \bibinfo {author} {\bibfnamefont
  {Z.}~\bibnamefont {Chen}}, \bibinfo {author} {\bibfnamefont {B.}~\bibnamefont
  {Chiaro}}, \bibinfo {author} {\bibfnamefont {A.}~\bibnamefont {Dunsworth}},
  \bibinfo {author} {\bibfnamefont {E.}~\bibnamefont {Jeffrey}}, \bibinfo
  {author} {\bibfnamefont {A.}~\bibnamefont {Megrant}}, \bibinfo {author}
  {\bibfnamefont {J.~Y.}\ \bibnamefont {Mutus}}, \bibinfo {author}
  {\bibfnamefont {P.~J.~J.}\ \bibnamefont {O'Malley}}, \bibinfo {author}
  {\bibfnamefont {C.~M.}\ \bibnamefont {Quintana}}, \bibinfo {author}
  {\bibfnamefont {D.}~\bibnamefont {Sank}}, \bibinfo {author} {\bibfnamefont
  {A.}~\bibnamefont {Vainsencher}}, \bibinfo {author} {\bibfnamefont
  {J.}~\bibnamefont {Wenner}}, \bibinfo {author} {\bibfnamefont {T.~C.}\
  \bibnamefont {White}}, \bibinfo {author} {\bibfnamefont {M.~R.}\ \bibnamefont
  {Geller}}, \bibinfo {author} {\bibfnamefont {A.~N.}\ \bibnamefont {Cleland}},
  \ and\ \bibinfo {author} {\bibfnamefont {J.~M.}\ \bibnamefont {Martinis}},\
  }\href {\doibase 10.1103/PhysRevLett.113.220502} {\bibfield  {journal}
  {\bibinfo  {journal} {Phys. Rev. Lett.}\ }\textbf {\bibinfo {volume} {113}},\
  \bibinfo {pages} {220502} (\bibinfo {year} {2014})}\BibitemShut {NoStop}%
\bibitem [{\citenamefont {Kelly}\ \emph {et~al.}(2015)\citenamefont {Kelly},
  \citenamefont {Barends}, \citenamefont {Fowler}, \citenamefont {Megrant},
  \citenamefont {Jeffrey}, \citenamefont {White}, \citenamefont {Sank},
  \citenamefont {Mutus}, \citenamefont {Campbell}, \citenamefont {Chen},
  \citenamefont {Chen}, \citenamefont {Chiaro}, \citenamefont {Dunsworth},
  \citenamefont {Hoi}, \citenamefont {Neill}, \citenamefont {O'Malley},
  \citenamefont {Quintana}, \citenamefont {Roushan}, \citenamefont
  {Vainsencher}, \citenamefont {Wenner}, \citenamefont {Cleland},\ and\
  \citenamefont {Martinis}}]{Kelly:2015aa}%
  \BibitemOpen
  \bibfield  {author} {\bibinfo {author} {\bibfnamefont {J.}~\bibnamefont
  {Kelly}}, \bibinfo {author} {\bibfnamefont {R.}~\bibnamefont {Barends}},
  \bibinfo {author} {\bibfnamefont {A.~G.}\ \bibnamefont {Fowler}}, \bibinfo
  {author} {\bibfnamefont {A.}~\bibnamefont {Megrant}}, \bibinfo {author}
  {\bibfnamefont {E.}~\bibnamefont {Jeffrey}}, \bibinfo {author} {\bibfnamefont
  {T.~C.}\ \bibnamefont {White}}, \bibinfo {author} {\bibfnamefont
  {D.}~\bibnamefont {Sank}}, \bibinfo {author} {\bibfnamefont {J.~Y.}\
  \bibnamefont {Mutus}}, \bibinfo {author} {\bibfnamefont {B.}~\bibnamefont
  {Campbell}}, \bibinfo {author} {\bibfnamefont {Y.}~\bibnamefont {Chen}},
  \bibinfo {author} {\bibfnamefont {Z.}~\bibnamefont {Chen}}, \bibinfo {author}
  {\bibfnamefont {B.}~\bibnamefont {Chiaro}}, \bibinfo {author} {\bibfnamefont
  {A.}~\bibnamefont {Dunsworth}}, \bibinfo {author} {\bibfnamefont {I.~C.}\
  \bibnamefont {Hoi}}, \bibinfo {author} {\bibfnamefont {C.}~\bibnamefont
  {Neill}}, \bibinfo {author} {\bibfnamefont {P.~J.~J.}\ \bibnamefont
  {O'Malley}}, \bibinfo {author} {\bibfnamefont {C.}~\bibnamefont {Quintana}},
  \bibinfo {author} {\bibfnamefont {P.}~\bibnamefont {Roushan}}, \bibinfo
  {author} {\bibfnamefont {A.}~\bibnamefont {Vainsencher}}, \bibinfo {author}
  {\bibfnamefont {J.}~\bibnamefont {Wenner}}, \bibinfo {author} {\bibfnamefont
  {A.~N.}\ \bibnamefont {Cleland}}, \ and\ \bibinfo {author} {\bibfnamefont
  {J.~M.}\ \bibnamefont {Martinis}},\ }\href {\doibase 10.1038/nature14270}
  {\bibfield  {journal} {\bibinfo  {journal} {Nature}\ }\textbf {\bibinfo
  {volume} {519}},\ \bibinfo {pages} {66} (\bibinfo {year} {2015})}\BibitemShut
  {NoStop}%
\bibitem [{\citenamefont {Arute}\ \emph {et~al.}(2019)\citenamefont {Arute},
  \citenamefont {Arya}, \citenamefont {Babbush}, \citenamefont {Bacon},
  \citenamefont {Bardin} \emph {et~al.}}]{Arute:2019aa}%
  \BibitemOpen
  \bibfield  {author} {\bibinfo {author} {\bibfnamefont {F.}~\bibnamefont
  {Arute}}, \bibinfo {author} {\bibfnamefont {K.}~\bibnamefont {Arya}},
  \bibinfo {author} {\bibfnamefont {R.}~\bibnamefont {Babbush}}, \bibinfo
  {author} {\bibfnamefont {D.}~\bibnamefont {Bacon}}, \bibinfo {author}
  {\bibfnamefont {J.~C.}\ \bibnamefont {Bardin}},  \emph {et~al.},\ }\href
  {\doibase 10.1038/s41586-019-1666-5} {\bibfield  {journal} {\bibinfo
  {journal} {Nature}\ }\textbf {\bibinfo {volume} {574}},\ \bibinfo {pages}
  {505} (\bibinfo {year} {2019})}\BibitemShut {NoStop}%
\bibitem [{\citenamefont {Reagor}\ \emph {et~al.}(2018)\citenamefont {Reagor},
  \citenamefont {Osborn}, \citenamefont {Tezak}, \citenamefont {Staley},
  \citenamefont {Prawiroatmodjo} \emph {et~al.}}]{Reagor:2018}%
  \BibitemOpen
  \bibfield  {author} {\bibinfo {author} {\bibfnamefont {M.}~\bibnamefont
  {Reagor}}, \bibinfo {author} {\bibfnamefont {C.~B.}\ \bibnamefont {Osborn}},
  \bibinfo {author} {\bibfnamefont {N.}~\bibnamefont {Tezak}}, \bibinfo
  {author} {\bibfnamefont {A.}~\bibnamefont {Staley}}, \bibinfo {author}
  {\bibfnamefont {G.}~\bibnamefont {Prawiroatmodjo}},  \emph {et~al.},\
  }\href@noop {} {\bibfield  {journal} {\bibinfo  {journal} {Science Advances}\
  }\textbf {\bibinfo {volume} {4}},\ \bibinfo {pages} {eaao3603} (\bibinfo
  {year} {2018})}\BibitemShut {NoStop}%
\bibitem [{\citenamefont {Debnath}\ \emph {et~al.}(2016)\citenamefont
  {Debnath}, \citenamefont {Linke}, \citenamefont {Figgatt}, \citenamefont
  {Landsman}, \citenamefont {Wright},\ and\ \citenamefont
  {Monroe}}]{Debnath:2016aa}%
  \BibitemOpen
  \bibfield  {author} {\bibinfo {author} {\bibfnamefont {S.}~\bibnamefont
  {Debnath}}, \bibinfo {author} {\bibfnamefont {N.~M.}\ \bibnamefont {Linke}},
  \bibinfo {author} {\bibfnamefont {C.}~\bibnamefont {Figgatt}}, \bibinfo
  {author} {\bibfnamefont {K.~A.}\ \bibnamefont {Landsman}}, \bibinfo {author}
  {\bibfnamefont {K.}~\bibnamefont {Wright}}, \ and\ \bibinfo {author}
  {\bibfnamefont {C.}~\bibnamefont {Monroe}},\ }\href@noop {} {\bibfield
  {journal} {\bibinfo  {journal} {Nature}\ }\textbf {\bibinfo {volume} {536}},\
  \bibinfo {pages} {63} (\bibinfo {year} {2016})}\BibitemShut {NoStop}%
\bibitem [{\citenamefont {Wright}\ \emph {et~al.}(2019)\citenamefont {Wright},
  \citenamefont {Beck}, \citenamefont {Debnath}, \citenamefont {Amini},
  \citenamefont {Nam} \emph {et~al.}}]{Wright:2019aa}%
  \BibitemOpen
  \bibfield  {author} {\bibinfo {author} {\bibfnamefont {K.}~\bibnamefont
  {Wright}}, \bibinfo {author} {\bibfnamefont {K.~M.}\ \bibnamefont {Beck}},
  \bibinfo {author} {\bibfnamefont {S.}~\bibnamefont {Debnath}}, \bibinfo
  {author} {\bibfnamefont {J.~M.}\ \bibnamefont {Amini}}, \bibinfo {author}
  {\bibfnamefont {Y.}~\bibnamefont {Nam}},  \emph {et~al.},\ }\href@noop {}
  {\bibfield  {journal} {\bibinfo  {journal} {Nature Communications}\ }\textbf
  {\bibinfo {volume} {10}},\ \bibinfo {pages} {5464} (\bibinfo {year}
  {2019})}\BibitemShut {NoStop}%
\bibitem [{\citenamefont {Preskill}(2018)}]{Preskill:2018}%
  \BibitemOpen
  \bibfield  {author} {\bibinfo {author} {\bibfnamefont {J.}~\bibnamefont
  {Preskill}},\ }\href@noop {} {\bibfield  {journal} {\bibinfo  {journal}
  {Quantum}\ }\textbf {\bibinfo {volume} {2}},\ \bibinfo {pages} {79} (\bibinfo
  {year} {2018})}\BibitemShut {NoStop}%
\bibitem [{\citenamefont {Bernien}\ \emph {et~al.}(2017)\citenamefont
  {Bernien}, \citenamefont {Schwartz}, \citenamefont {Keesling}, \citenamefont
  {Levine}, \citenamefont {Omran}, \citenamefont {Pichler}, \citenamefont
  {Choi}, \citenamefont {Zibrov}, \citenamefont {Endres}, \citenamefont
  {Greiner}, \citenamefont {Vuleti{\'c}},\ and\ \citenamefont
  {Lukin}}]{Bernien:2017aa}%
  \BibitemOpen
  \bibfield  {author} {\bibinfo {author} {\bibfnamefont {H.}~\bibnamefont
  {Bernien}}, \bibinfo {author} {\bibfnamefont {S.}~\bibnamefont {Schwartz}},
  \bibinfo {author} {\bibfnamefont {A.}~\bibnamefont {Keesling}}, \bibinfo
  {author} {\bibfnamefont {H.}~\bibnamefont {Levine}}, \bibinfo {author}
  {\bibfnamefont {A.}~\bibnamefont {Omran}}, \bibinfo {author} {\bibfnamefont
  {H.}~\bibnamefont {Pichler}}, \bibinfo {author} {\bibfnamefont
  {S.}~\bibnamefont {Choi}}, \bibinfo {author} {\bibfnamefont {A.~S.}\
  \bibnamefont {Zibrov}}, \bibinfo {author} {\bibfnamefont {M.}~\bibnamefont
  {Endres}}, \bibinfo {author} {\bibfnamefont {M.}~\bibnamefont {Greiner}},
  \bibinfo {author} {\bibfnamefont {V.}~\bibnamefont {Vuleti{\'c}}}, \ and\
  \bibinfo {author} {\bibfnamefont {M.~D.}\ \bibnamefont {Lukin}},\ }\href@noop
  {} {\bibfield  {journal} {\bibinfo  {journal} {Nature}\ }\textbf {\bibinfo
  {volume} {551}},\ \bibinfo {pages} {579} (\bibinfo {year}
  {2017})}\BibitemShut {NoStop}%
\bibitem [{\citenamefont {Nam}\ and\ \citenamefont
  {Maslov}(2019)}]{Nam:2019aa}%
  \BibitemOpen
  \bibfield  {author} {\bibinfo {author} {\bibfnamefont {Y.}~\bibnamefont
  {Nam}}\ and\ \bibinfo {author} {\bibfnamefont {D.}~\bibnamefont {Maslov}},\
  }\href@noop {} {\bibfield  {journal} {\bibinfo  {journal} {npj Quantum
  Information}\ }\textbf {\bibinfo {volume} {5}},\ \bibinfo {pages} {44}
  (\bibinfo {year} {2019})}\BibitemShut {NoStop}%
\bibitem [{\citenamefont {Nersisyan}\ \emph {et~al.}(2019)\citenamefont
  {Nersisyan}, \citenamefont {Poletto}, \citenamefont {Alidoust}, \citenamefont
  {Manenti}, \citenamefont {Renzas}, \citenamefont {Bui}, \citenamefont {Vu},
  \citenamefont {Whyland}, \citenamefont {Mohan}, \citenamefont {Sete},
  \citenamefont {Stanwyck}, \citenamefont {Bestwick},\ and\ \citenamefont
  {Reagor}}]{Nersisyan:2019}%
  \BibitemOpen
  \bibfield  {author} {\bibinfo {author} {\bibfnamefont {A.}~\bibnamefont
  {Nersisyan}}, \bibinfo {author} {\bibfnamefont {S.}~\bibnamefont {Poletto}},
  \bibinfo {author} {\bibfnamefont {N.}~\bibnamefont {Alidoust}}, \bibinfo
  {author} {\bibfnamefont {R.}~\bibnamefont {Manenti}}, \bibinfo {author}
  {\bibfnamefont {R.}~\bibnamefont {Renzas}}, \bibinfo {author} {\bibfnamefont
  {C.}~\bibnamefont {Bui}}, \bibinfo {author} {\bibfnamefont {K.}~\bibnamefont
  {Vu}}, \bibinfo {author} {\bibfnamefont {T.}~\bibnamefont {Whyland}},
  \bibinfo {author} {\bibfnamefont {Y.}~\bibnamefont {Mohan}}, \bibinfo
  {author} {\bibfnamefont {E.~A.}\ \bibnamefont {Sete}}, \bibinfo {author}
  {\bibfnamefont {S.}~\bibnamefont {Stanwyck}}, \bibinfo {author}
  {\bibfnamefont {A.}~\bibnamefont {Bestwick}}, \ and\ \bibinfo {author}
  {\bibfnamefont {M.}~\bibnamefont {Reagor}},\ }in\ \href@noop {} {\emph
  {\bibinfo {booktitle} {2019 IEEE International Electron Devices Meeting
  (IEDM)}}}\ (\bibinfo {year} {2019})\ pp.\ \bibinfo {pages}
  {31.1.1--31.1.4}\BibitemShut {NoStop}%
\bibitem [{\citenamefont {Nguyen}\ \emph {et~al.}(2019)\citenamefont {Nguyen},
  \citenamefont {Lin}, \citenamefont {Somoroff}, \citenamefont {Mencia},
  \citenamefont {Grabon},\ and\ \citenamefont {Manucharyan}}]{Nguyen:2019}%
  \BibitemOpen
  \bibfield  {author} {\bibinfo {author} {\bibfnamefont {L.~B.}\ \bibnamefont
  {Nguyen}}, \bibinfo {author} {\bibfnamefont {Y.-H.}\ \bibnamefont {Lin}},
  \bibinfo {author} {\bibfnamefont {A.}~\bibnamefont {Somoroff}}, \bibinfo
  {author} {\bibfnamefont {R.}~\bibnamefont {Mencia}}, \bibinfo {author}
  {\bibfnamefont {N.}~\bibnamefont {Grabon}}, \ and\ \bibinfo {author}
  {\bibfnamefont {V.~E.}\ \bibnamefont {Manucharyan}},\ }\href@noop {}
  {\bibfield  {journal} {\bibinfo  {journal} {Physical Review X}\ }\textbf
  {\bibinfo {volume} {9}},\ \bibinfo {pages} {041041} (\bibinfo {year}
  {2019})}\BibitemShut {NoStop}%
\bibitem [{\citenamefont {Place}\ \emph {et~al.}()\citenamefont {Place},
  \citenamefont {Rodgers}, \citenamefont {Mundata}, \citenamefont {Smitham},
  \citenamefont {Fitzpatrick} \emph {et~al.}}]{Alex:2020}%
  \BibitemOpen
  \bibfield  {author} {\bibinfo {author} {\bibfnamefont {A.~P.~M.}\
  \bibnamefont {Place}}, \bibinfo {author} {\bibfnamefont {L.~V.~H.}\
  \bibnamefont {Rodgers}}, \bibinfo {author} {\bibfnamefont {P.}~\bibnamefont
  {Mundata}}, \bibinfo {author} {\bibfnamefont {B.~M.}\ \bibnamefont
  {Smitham}}, \bibinfo {author} {\bibfnamefont {M.}~\bibnamefont
  {Fitzpatrick}},  \emph {et~al.},\ }\href@noop {} {\bibinfo  {journal}
  {arXiv:2003.00024v1}\ }\BibitemShut {NoStop}%
\bibitem [{\citenamefont {Khaneja}\ \emph {et~al.}(2005)\citenamefont
  {Khaneja}, \citenamefont {Reiss}, \citenamefont {Kehlet}, \citenamefont
  {Schulte-Herbruggen},\ and\ \citenamefont {Glaser}}]{Khaneja:2005}%
  \BibitemOpen
\bibfield  {journal} {  }\bibfield  {author} {\bibinfo {author} {\bibfnamefont
  {N.}~\bibnamefont {Khaneja}}, \bibinfo {author} {\bibfnamefont
  {T.}~\bibnamefont {Reiss}}, \bibinfo {author} {\bibfnamefont
  {C.}~\bibnamefont {Kehlet}}, \bibinfo {author} {\bibfnamefont
  {T.}~\bibnamefont {Schulte-Herbruggen}}, \ and\ \bibinfo {author}
  {\bibfnamefont {S.~J.}\ \bibnamefont {Glaser}},\ }\href@noop {} {\bibfield
  {journal} {\bibinfo  {journal} {Journal of Magnetic Resonance}\ }\textbf
  {\bibinfo {volume} {172}},\ \bibinfo {pages} {296} (\bibinfo {year}
  {2005})}\BibitemShut {NoStop}%
\bibitem [{\citenamefont {Floether}\ \emph {et~al.}(2012)\citenamefont
  {Floether}, \citenamefont {de~Fouquieres},\ and\ \citenamefont
  {Shirmer}}]{Floether:2012}%
  \BibitemOpen
  \bibfield  {author} {\bibinfo {author} {\bibfnamefont {F.~F.}\ \bibnamefont
  {Floether}}, \bibinfo {author} {\bibfnamefont {P.}~\bibnamefont
  {de~Fouquieres}}, \ and\ \bibinfo {author} {\bibfnamefont {S.~G.}\
  \bibnamefont {Shirmer}},\ }\href@noop {} {\bibfield  {journal} {\bibinfo
  {journal} {New Journal of Physics}\ }\textbf {\bibinfo {volume} {14}},\
  \bibinfo {pages} {073023} (\bibinfo {year} {2012})}\BibitemShut {NoStop}%
\bibitem [{\citenamefont {Holland}\ \emph {et~al.}()\citenamefont {Holland},
  \citenamefont {Wendt}, \citenamefont {Kravvaris}, \citenamefont {Wu},
  \citenamefont {Ormand}, \citenamefont {L~DuBois}, \citenamefont {Quaglioni},\
  and\ \citenamefont {Pederiva}}]{Holland:2019}%
  \BibitemOpen
  \bibfield  {author} {\bibinfo {author} {\bibfnamefont {E.~T.}\ \bibnamefont
  {Holland}}, \bibinfo {author} {\bibfnamefont {K.~A.}\ \bibnamefont {Wendt}},
  \bibinfo {author} {\bibfnamefont {K.}~\bibnamefont {Kravvaris}}, \bibinfo
  {author} {\bibfnamefont {X.}~\bibnamefont {Wu}}, \bibinfo {author}
  {\bibfnamefont {E.~W.}\ \bibnamefont {Ormand}}, \bibinfo {author}
  {\bibfnamefont {J.}~\bibnamefont {L~DuBois}}, \bibinfo {author}
  {\bibfnamefont {S.}~\bibnamefont {Quaglioni}}, \ and\ \bibinfo {author}
  {\bibfnamefont {F.}~\bibnamefont {Pederiva}},\ }\href@noop {} {\bibinfo
  {journal} {arXiv:1908.08222v1}\ }\BibitemShut {NoStop}%
\bibitem [{\citenamefont {Machnes}\ \emph {et~al.}(2011)\citenamefont
  {Machnes}, \citenamefont {Sander}, \citenamefont {Glaser}, \citenamefont
  {de~Fouqui\`eres}, \citenamefont {Gruslys}, \citenamefont {Schirmer},\ and\
  \citenamefont {Schulte-Herbr\"uggen}}]{Machnes:2011}%
  \BibitemOpen
\bibfield  {journal} {  }\bibfield  {author} {\bibinfo {author} {\bibfnamefont
  {S.}~\bibnamefont {Machnes}}, \bibinfo {author} {\bibfnamefont
  {U.}~\bibnamefont {Sander}}, \bibinfo {author} {\bibfnamefont {S.~J.}\
  \bibnamefont {Glaser}}, \bibinfo {author} {\bibfnamefont {P.}~\bibnamefont
  {de~Fouqui\`eres}}, \bibinfo {author} {\bibfnamefont {A.}~\bibnamefont
  {Gruslys}}, \bibinfo {author} {\bibfnamefont {S.}~\bibnamefont {Schirmer}}, \
  and\ \bibinfo {author} {\bibfnamefont {T.}~\bibnamefont
  {Schulte-Herbr\"uggen}},\ }\href@noop {} {\bibfield  {journal} {\bibinfo
  {journal} {Physical Review A}\ }\textbf {\bibinfo {volume} {84}},\ \bibinfo
  {pages} {022305} (\bibinfo {year} {2011})}\BibitemShut {NoStop}%
\bibitem [{\citenamefont {Cheng}\ \emph {et~al.}()\citenamefont {Cheng},
  \citenamefont {Deng},\ and\ \citenamefont {Qian}}]{Cheng:2020}%
  \BibitemOpen
  \bibfield  {author} {\bibinfo {author} {\bibfnamefont {J.}~\bibnamefont
  {Cheng}}, \bibinfo {author} {\bibfnamefont {H.}~\bibnamefont {Deng}}, \ and\
  \bibinfo {author} {\bibfnamefont {X.}~\bibnamefont {Qian}},\ }\href@noop {}
  {\bibinfo  {journal} {arXiv:2003.00376v1}\ }\BibitemShut {NoStop}%
\bibitem [{\citenamefont {Paraoanu}(2014)}]{Paraoanu}%
  \BibitemOpen
\bibfield  {journal} {  }\bibfield  {author} {\bibinfo {author} {\bibfnamefont
  {G.~S.}\ \bibnamefont {Paraoanu}},\ }\href@noop {} {\bibfield  {journal}
  {\bibinfo  {journal} {Journal of Low Temperature Physics}\ }\textbf {\bibinfo
  {volume} {175}},\ \bibinfo {pages} {633} (\bibinfo {year}
  {2014})}\BibitemShut {NoStop}%
\bibitem [{\citenamefont {Barends}\ \emph {et~al.}(2014)\citenamefont
  {Barends}, \citenamefont {Kelly}, \citenamefont {Megrant}, \citenamefont
  {Veitia}, \citenamefont {Sank} \emph {et~al.}}]{Barends:2014aa}%
  \BibitemOpen
  \bibfield  {author} {\bibinfo {author} {\bibfnamefont {R.}~\bibnamefont
  {Barends}}, \bibinfo {author} {\bibfnamefont {J.}~\bibnamefont {Kelly}},
  \bibinfo {author} {\bibfnamefont {A.}~\bibnamefont {Megrant}}, \bibinfo
  {author} {\bibfnamefont {A.}~\bibnamefont {Veitia}}, \bibinfo {author}
  {\bibfnamefont {D.}~\bibnamefont {Sank}},  \emph {et~al.},\ }\href@noop {}
  {\bibfield  {journal} {\bibinfo  {journal} {Nature}\ }\textbf {\bibinfo
  {volume} {508}},\ \bibinfo {pages} {500} (\bibinfo {year}
  {2014})}\BibitemShut {NoStop}%
\bibitem [{\citenamefont {Sheldon}\ \emph {et~al.}(2016)\citenamefont
  {Sheldon}, \citenamefont {Bishop}, \citenamefont {Magesan}, \citenamefont
  {Filipp}, \citenamefont {Chow},\ and\ \citenamefont
  {Gambetta}}]{Sheldon:2016}%
  \BibitemOpen
  \bibfield  {author} {\bibinfo {author} {\bibfnamefont {S.}~\bibnamefont
  {Sheldon}}, \bibinfo {author} {\bibfnamefont {L.~S.}\ \bibnamefont {Bishop}},
  \bibinfo {author} {\bibfnamefont {E.}~\bibnamefont {Magesan}}, \bibinfo
  {author} {\bibfnamefont {S.}~\bibnamefont {Filipp}}, \bibinfo {author}
  {\bibfnamefont {J.~M.}\ \bibnamefont {Chow}}, \ and\ \bibinfo {author}
  {\bibfnamefont {J.~M.}\ \bibnamefont {Gambetta}},\ }\href@noop {} {\bibfield
  {journal} {\bibinfo  {journal} {Physical Review A}\ }\textbf {\bibinfo
  {volume} {93}},\ \bibinfo {pages} {012301} (\bibinfo {year}
  {2016})}\BibitemShut {NoStop}%
\bibitem [{\citenamefont {Rol}\ \emph {et~al.}(2017)\citenamefont {Rol},
  \citenamefont {Bultink}, \citenamefont {O'Brien}, \citenamefont {de~Jong},
  \citenamefont {Theis}, \citenamefont {Fu}, \citenamefont {Luthi},
  \citenamefont {Vermeulen}, \citenamefont {de~Sterke}, \citenamefont {Bruno},
  \citenamefont {Deurloo}, \citenamefont {Schouten}, \citenamefont {Wilhelm},\
  and\ \citenamefont {DiCarlo}}]{Rol:2017}%
  \BibitemOpen
  \bibfield  {author} {\bibinfo {author} {\bibfnamefont {M.~A.}\ \bibnamefont
  {Rol}}, \bibinfo {author} {\bibfnamefont {C.~C.}\ \bibnamefont {Bultink}},
  \bibinfo {author} {\bibfnamefont {T.~E.}\ \bibnamefont {O'Brien}}, \bibinfo
  {author} {\bibfnamefont {S.~R.}\ \bibnamefont {de~Jong}}, \bibinfo {author}
  {\bibfnamefont {L.~S.}\ \bibnamefont {Theis}}, \bibinfo {author}
  {\bibfnamefont {X.}~\bibnamefont {Fu}}, \bibinfo {author} {\bibfnamefont
  {F.}~\bibnamefont {Luthi}}, \bibinfo {author} {\bibfnamefont {R.~F.~L.}\
  \bibnamefont {Vermeulen}}, \bibinfo {author} {\bibfnamefont {J.~C.}\
  \bibnamefont {de~Sterke}}, \bibinfo {author} {\bibfnamefont {A.}~\bibnamefont
  {Bruno}}, \bibinfo {author} {\bibfnamefont {D.}~\bibnamefont {Deurloo}},
  \bibinfo {author} {\bibfnamefont {R.~N.}\ \bibnamefont {Schouten}}, \bibinfo
  {author} {\bibfnamefont {F.~K.}\ \bibnamefont {Wilhelm}}, \ and\ \bibinfo
  {author} {\bibfnamefont {L.}~\bibnamefont {DiCarlo}},\ }\href@noop {}
  {\bibfield  {journal} {\bibinfo  {journal} {Physical Review Applied}\
  }\textbf {\bibinfo {volume} {7}},\ \bibinfo {pages} {041001} (\bibinfo {year}
  {2017})}\BibitemShut {NoStop}%
\bibitem [{\citenamefont {Dewes}\ \emph {et~al.}(2012)\citenamefont {Dewes},
  \citenamefont {Ong}, \citenamefont {Schmitt}, \citenamefont {Lauro},
  \citenamefont {Boulant}, \citenamefont {Bertet}, \citenamefont {Vion},\ and\
  \citenamefont {Esteve}}]{Dewes:2012}%
  \BibitemOpen
  \bibfield  {author} {\bibinfo {author} {\bibfnamefont {A.}~\bibnamefont
  {Dewes}}, \bibinfo {author} {\bibfnamefont {F.~R.}\ \bibnamefont {Ong}},
  \bibinfo {author} {\bibfnamefont {V.}~\bibnamefont {Schmitt}}, \bibinfo
  {author} {\bibfnamefont {R.}~\bibnamefont {Lauro}}, \bibinfo {author}
  {\bibfnamefont {N.}~\bibnamefont {Boulant}}, \bibinfo {author} {\bibfnamefont
  {P.}~\bibnamefont {Bertet}}, \bibinfo {author} {\bibfnamefont
  {D.}~\bibnamefont {Vion}}, \ and\ \bibinfo {author} {\bibfnamefont
  {D.}~\bibnamefont {Esteve}},\ }\href@noop {} {\bibfield  {journal} {\bibinfo
  {journal} {Physical Review Letters}\ }\textbf {\bibinfo {volume} {108}},\
  \bibinfo {pages} {057002} (\bibinfo {year} {2012})}\BibitemShut {NoStop}%
\bibitem [{\citenamefont {Rosenblum}\ \emph
  {et~al.}(2018{\natexlab{a}})\citenamefont {Rosenblum}, \citenamefont {Gao},
  \citenamefont {Reinhold}, \citenamefont {Wang}, \citenamefont {Axline},
  \citenamefont {Frunzio}, \citenamefont {Girvin}, \citenamefont {Jiang},
  \citenamefont {Mirrahimi}, \citenamefont {Devoret},\ and\ \citenamefont
  {Schoelkopf}}]{Rosenblum:2018aa}%
  \BibitemOpen
  \bibfield  {author} {\bibinfo {author} {\bibfnamefont {S.}~\bibnamefont
  {Rosenblum}}, \bibinfo {author} {\bibfnamefont {Y.~Y.}\ \bibnamefont {Gao}},
  \bibinfo {author} {\bibfnamefont {P.}~\bibnamefont {Reinhold}}, \bibinfo
  {author} {\bibfnamefont {C.}~\bibnamefont {Wang}}, \bibinfo {author}
  {\bibfnamefont {C.~J.}\ \bibnamefont {Axline}}, \bibinfo {author}
  {\bibfnamefont {L.}~\bibnamefont {Frunzio}}, \bibinfo {author} {\bibfnamefont
  {S.~M.}\ \bibnamefont {Girvin}}, \bibinfo {author} {\bibfnamefont
  {L.}~\bibnamefont {Jiang}}, \bibinfo {author} {\bibfnamefont
  {M.}~\bibnamefont {Mirrahimi}}, \bibinfo {author} {\bibfnamefont {M.~H.}\
  \bibnamefont {Devoret}}, \ and\ \bibinfo {author} {\bibfnamefont {R.~J.}\
  \bibnamefont {Schoelkopf}},\ }\href@noop {} {\bibfield  {journal} {\bibinfo
  {journal} {Nature Communications}\ }\textbf {\bibinfo {volume} {9}},\
  \bibinfo {pages} {652} (\bibinfo {year} {2018}{\natexlab{a}})}\BibitemShut
  {NoStop}%
\bibitem [{\citenamefont {Hong}\ \emph {et~al.}(2020)\citenamefont {Hong},
  \citenamefont {Papageorge}, \citenamefont {Sivarajah}, \citenamefont
  {Crossman}, \citenamefont {Didier}, \citenamefont {Polloreno}, \citenamefont
  {Sete}, \citenamefont {Turkowski}, \citenamefont {da~Silva},\ and\
  \citenamefont {Johnson}}]{Hong:2020}%
  \BibitemOpen
  \bibfield  {author} {\bibinfo {author} {\bibfnamefont {S.~S.}\ \bibnamefont
  {Hong}}, \bibinfo {author} {\bibfnamefont {A.~T.}\ \bibnamefont
  {Papageorge}}, \bibinfo {author} {\bibfnamefont {P.}~\bibnamefont
  {Sivarajah}}, \bibinfo {author} {\bibfnamefont {G.}~\bibnamefont {Crossman}},
  \bibinfo {author} {\bibfnamefont {N.}~\bibnamefont {Didier}}, \bibinfo
  {author} {\bibfnamefont {A.~M.}\ \bibnamefont {Polloreno}}, \bibinfo {author}
  {\bibfnamefont {E.~A.}\ \bibnamefont {Sete}}, \bibinfo {author}
  {\bibfnamefont {S.~W.}\ \bibnamefont {Turkowski}}, \bibinfo {author}
  {\bibfnamefont {M.~P.}\ \bibnamefont {da~Silva}}, \ and\ \bibinfo {author}
  {\bibfnamefont {B.~R.}\ \bibnamefont {Johnson}},\ }\href@noop {} {\bibfield
  {journal} {\bibinfo  {journal} {Physical Review A}\ }\textbf {\bibinfo
  {volume} {101}},\ \bibinfo {pages} {012302} (\bibinfo {year}
  {2020})}\BibitemShut {NoStop}%
\bibitem [{\citenamefont {Kjaergaard}\ \emph {et~al.}(2020)\citenamefont
  {Kjaergaard}, \citenamefont {Schwartz}, \citenamefont {Greene}, \citenamefont
  {Samach}, \citenamefont {Bengtsson} \emph {et~al.}}]{Kjaergaard:2020b}%
  \BibitemOpen
  \bibfield  {author} {\bibinfo {author} {\bibfnamefont {M.}~\bibnamefont
  {Kjaergaard}}, \bibinfo {author} {\bibfnamefont {M.~E.}\ \bibnamefont
  {Schwartz}}, \bibinfo {author} {\bibfnamefont {A.}~\bibnamefont {Greene}},
  \bibinfo {author} {\bibfnamefont {G.~O.}\ \bibnamefont {Samach}}, \bibinfo
  {author} {\bibfnamefont {A.}~\bibnamefont {Bengtsson}},  \emph {et~al.},\
  }\href@noop {} {\bibfield  {journal} {\bibinfo  {journal}
  {arXiv:2001.08838v2}\ } (\bibinfo {year} {2020})}\BibitemShut {NoStop}%
\bibitem [{\citenamefont {Lanyon}\ \emph {et~al.}(2011)\citenamefont {Lanyon},
  \citenamefont {Hempel}, \citenamefont {Nigg}, \citenamefont {M{\"u}ller},
  \citenamefont {Gerritsma}, \citenamefont {Z{\"a}hringer}, \citenamefont
  {Schindler}, \citenamefont {Barreiro}, \citenamefont {Rambach}, \citenamefont
  {Kirchmair}, \citenamefont {Hennrich}, \citenamefont {Zoller}, \citenamefont
  {Blatt},\ and\ \citenamefont {Roos}}]{Lanyon57}%
  \BibitemOpen
  \bibfield  {author} {\bibinfo {author} {\bibfnamefont {B.~P.}\ \bibnamefont
  {Lanyon}}, \bibinfo {author} {\bibfnamefont {C.}~\bibnamefont {Hempel}},
  \bibinfo {author} {\bibfnamefont {D.}~\bibnamefont {Nigg}}, \bibinfo {author}
  {\bibfnamefont {M.}~\bibnamefont {M{\"u}ller}}, \bibinfo {author}
  {\bibfnamefont {R.}~\bibnamefont {Gerritsma}}, \bibinfo {author}
  {\bibfnamefont {F.}~\bibnamefont {Z{\"a}hringer}}, \bibinfo {author}
  {\bibfnamefont {P.}~\bibnamefont {Schindler}}, \bibinfo {author}
  {\bibfnamefont {J.~T.}\ \bibnamefont {Barreiro}}, \bibinfo {author}
  {\bibfnamefont {M.}~\bibnamefont {Rambach}}, \bibinfo {author} {\bibfnamefont
  {G.}~\bibnamefont {Kirchmair}}, \bibinfo {author} {\bibfnamefont
  {M.}~\bibnamefont {Hennrich}}, \bibinfo {author} {\bibfnamefont
  {P.}~\bibnamefont {Zoller}}, \bibinfo {author} {\bibfnamefont
  {R.}~\bibnamefont {Blatt}}, \ and\ \bibinfo {author} {\bibfnamefont {C.~F.}\
  \bibnamefont {Roos}},\ }\href@noop {} {\bibfield  {journal} {\bibinfo
  {journal} {Science}\ }\textbf {\bibinfo {volume} {334}},\ \bibinfo {pages}
  {57} (\bibinfo {year} {2011})}\BibitemShut {NoStop}%
\bibitem [{\citenamefont {Barends}\ \emph {et~al.}(2015)\citenamefont
  {Barends}, \citenamefont {Lamata}, \citenamefont {Kelly}, \citenamefont
  {Garcia-Alvarez}, \citenamefont {Fowler} \emph {et~al.}}]{Barends:2015}%
  \BibitemOpen
  \bibfield  {author} {\bibinfo {author} {\bibfnamefont {R.}~\bibnamefont
  {Barends}}, \bibinfo {author} {\bibfnamefont {L.}~\bibnamefont {Lamata}},
  \bibinfo {author} {\bibfnamefont {J.}~\bibnamefont {Kelly}}, \bibinfo
  {author} {\bibfnamefont {L.}~\bibnamefont {Garcia-Alvarez}}, \bibinfo
  {author} {\bibfnamefont {A.~G.}\ \bibnamefont {Fowler}},  \emph {et~al.},\
  }\href@noop {} {\bibfield  {journal} {\bibinfo  {journal} {Nature
  Communications}\ }\textbf {\bibinfo {volume} {6}},\ \bibinfo {pages} {7654}
  (\bibinfo {year} {2015})}\BibitemShut {NoStop}%
\bibitem [{\citenamefont {O'Malley}\ \emph {et~al.}(2016)\citenamefont
  {O'Malley}, \citenamefont {Babbush}, \citenamefont {Kivlichan}, \citenamefont
  {Romero}, \citenamefont {McClean}, \citenamefont {Barends}, \citenamefont
  {Kelly}, \citenamefont {Roushan}, \citenamefont {Tranter}, \citenamefont
  {Ding}, \citenamefont {Campbell}, \citenamefont {Chen}, \citenamefont {Chen},
  \citenamefont {Chiaro}, \citenamefont {Dunsworth}, \citenamefont {Fowler},
  \citenamefont {Jeffrey}, \citenamefont {Lucero}, \citenamefont {Megrant},
  \citenamefont {Mutus}, \citenamefont {Neeley}, \citenamefont {Neill},
  \citenamefont {Quintana}, \citenamefont {Sank}, \citenamefont {Vainsencher},
  \citenamefont {Wenner}, \citenamefont {White}, \citenamefont {Coveney},
  \citenamefont {Love}, \citenamefont {Neven}, \citenamefont {Aspuru-Guzik},\
  and\ \citenamefont {Martinis}}]{OMalley:2016}%
  \BibitemOpen
  \bibfield  {author} {\bibinfo {author} {\bibfnamefont {P.~J.~J.}\
  \bibnamefont {O'Malley}}, \bibinfo {author} {\bibfnamefont {R.}~\bibnamefont
  {Babbush}}, \bibinfo {author} {\bibfnamefont {I.~D.}\ \bibnamefont
  {Kivlichan}}, \bibinfo {author} {\bibfnamefont {J.}~\bibnamefont {Romero}},
  \bibinfo {author} {\bibfnamefont {J.~R.}\ \bibnamefont {McClean}}, \bibinfo
  {author} {\bibfnamefont {R.}~\bibnamefont {Barends}}, \bibinfo {author}
  {\bibfnamefont {J.}~\bibnamefont {Kelly}}, \bibinfo {author} {\bibfnamefont
  {P.}~\bibnamefont {Roushan}}, \bibinfo {author} {\bibfnamefont
  {A.}~\bibnamefont {Tranter}}, \bibinfo {author} {\bibfnamefont
  {N.}~\bibnamefont {Ding}}, \bibinfo {author} {\bibfnamefont {B.}~\bibnamefont
  {Campbell}}, \bibinfo {author} {\bibfnamefont {Y.}~\bibnamefont {Chen}},
  \bibinfo {author} {\bibfnamefont {Z.}~\bibnamefont {Chen}}, \bibinfo {author}
  {\bibfnamefont {B.}~\bibnamefont {Chiaro}}, \bibinfo {author} {\bibfnamefont
  {A.}~\bibnamefont {Dunsworth}}, \bibinfo {author} {\bibfnamefont {A.~G.}\
  \bibnamefont {Fowler}}, \bibinfo {author} {\bibfnamefont {E.}~\bibnamefont
  {Jeffrey}}, \bibinfo {author} {\bibfnamefont {E.}~\bibnamefont {Lucero}},
  \bibinfo {author} {\bibfnamefont {A.}~\bibnamefont {Megrant}}, \bibinfo
  {author} {\bibfnamefont {J.~Y.}\ \bibnamefont {Mutus}}, \bibinfo {author}
  {\bibfnamefont {M.}~\bibnamefont {Neeley}}, \bibinfo {author} {\bibfnamefont
  {C.}~\bibnamefont {Neill}}, \bibinfo {author} {\bibfnamefont
  {C.}~\bibnamefont {Quintana}}, \bibinfo {author} {\bibfnamefont
  {D.}~\bibnamefont {Sank}}, \bibinfo {author} {\bibfnamefont {A.}~\bibnamefont
  {Vainsencher}}, \bibinfo {author} {\bibfnamefont {J.}~\bibnamefont {Wenner}},
  \bibinfo {author} {\bibfnamefont {T.~C.}\ \bibnamefont {White}}, \bibinfo
  {author} {\bibfnamefont {P.~V.}\ \bibnamefont {Coveney}}, \bibinfo {author}
  {\bibfnamefont {P.~J.}\ \bibnamefont {Love}}, \bibinfo {author}
  {\bibfnamefont {H.}~\bibnamefont {Neven}}, \bibinfo {author} {\bibfnamefont
  {A.}~\bibnamefont {Aspuru-Guzik}}, \ and\ \bibinfo {author} {\bibfnamefont
  {J.~M.}\ \bibnamefont {Martinis}},\ }\href {\doibase
  10.1103/PhysRevX.6.031007} {\bibfield  {journal} {\bibinfo  {journal} {Phys.
  Rev. X}\ }\textbf {\bibinfo {volume} {6}},\ \bibinfo {pages} {031007}
  (\bibinfo {year} {2016})}\BibitemShut {NoStop}%
\bibitem [{\citenamefont {Kandala}\ \emph {et~al.}(2017)\citenamefont
  {Kandala}, \citenamefont {Mezzacapo}, \citenamefont {Temme}, \citenamefont
  {Takita}, \citenamefont {Brink}, \citenamefont {Chow},\ and\ \citenamefont
  {Gambetta}}]{Kandala:2017aa}%
  \BibitemOpen
  \bibfield  {author} {\bibinfo {author} {\bibfnamefont {A.}~\bibnamefont
  {Kandala}}, \bibinfo {author} {\bibfnamefont {A.}~\bibnamefont {Mezzacapo}},
  \bibinfo {author} {\bibfnamefont {K.}~\bibnamefont {Temme}}, \bibinfo
  {author} {\bibfnamefont {M.}~\bibnamefont {Takita}}, \bibinfo {author}
  {\bibfnamefont {M.}~\bibnamefont {Brink}}, \bibinfo {author} {\bibfnamefont
  {J.~M.}\ \bibnamefont {Chow}}, \ and\ \bibinfo {author} {\bibfnamefont
  {J.~M.}\ \bibnamefont {Gambetta}},\ }\href@noop {} {\bibfield  {journal}
  {\bibinfo  {journal} {Nature}\ }\textbf {\bibinfo {volume} {549}},\ \bibinfo
  {pages} {242} (\bibinfo {year} {2017})}\BibitemShut {NoStop}%
\bibitem [{\citenamefont {Shi}\ \emph {et~al.}()\citenamefont {Shi},
  \citenamefont {Castelli}, \citenamefont {Wu}, \citenamefont {Joseph},
  \citenamefont {Geyko}, \citenamefont {Graziani}, \citenamefont {Libby},
  \citenamefont {Parker}, \citenamefont {Rosen}, \citenamefont {Martinez},\
  and\ \citenamefont {LL~DuBois}}]{shi2020quantum}%
  \BibitemOpen
  \bibfield  {author} {\bibinfo {author} {\bibfnamefont {Y.}~\bibnamefont
  {Shi}}, \bibinfo {author} {\bibfnamefont {A.~R.}\ \bibnamefont {Castelli}},
  \bibinfo {author} {\bibfnamefont {X.}~\bibnamefont {Wu}}, \bibinfo {author}
  {\bibfnamefont {I.}~\bibnamefont {Joseph}}, \bibinfo {author} {\bibfnamefont
  {V.}~\bibnamefont {Geyko}}, \bibinfo {author} {\bibfnamefont {F.~R.}\
  \bibnamefont {Graziani}}, \bibinfo {author} {\bibfnamefont {S.~B.}\
  \bibnamefont {Libby}}, \bibinfo {author} {\bibfnamefont {J.~B.}\ \bibnamefont
  {Parker}}, \bibinfo {author} {\bibfnamefont {Y.~J.}\ \bibnamefont {Rosen}},
  \bibinfo {author} {\bibfnamefont {L.~A.}\ \bibnamefont {Martinez}}, \ and\
  \bibinfo {author} {\bibfnamefont {J.}~\bibnamefont {LL~DuBois}},\ }\href@noop
  {} {\bibinfo  {journal} {arXiv:2004.06885}\ }\BibitemShut {NoStop}%
\bibitem [{\citenamefont {Neeley}\ \emph {et~al.}(2009)\citenamefont {Neeley},
  \citenamefont {Ansmann}, \citenamefont {Bialczak}, \citenamefont {Hofheinz},
  \citenamefont {Lucero}, \citenamefont {O'Connell}, \citenamefont {Sank},
  \citenamefont {Wang}, \citenamefont {Wenner}, \citenamefont {Cleland},
  \citenamefont {Geller},\ and\ \citenamefont {Martinis}}]{Neeley:2009}%
  \BibitemOpen
\bibfield  {journal} {  }\bibfield  {author} {\bibinfo {author} {\bibfnamefont
  {M.}~\bibnamefont {Neeley}}, \bibinfo {author} {\bibfnamefont
  {M.}~\bibnamefont {Ansmann}}, \bibinfo {author} {\bibfnamefont {R.~C.}\
  \bibnamefont {Bialczak}}, \bibinfo {author} {\bibfnamefont {M.}~\bibnamefont
  {Hofheinz}}, \bibinfo {author} {\bibfnamefont {E.}~\bibnamefont {Lucero}},
  \bibinfo {author} {\bibfnamefont {A.~D.}\ \bibnamefont {O'Connell}}, \bibinfo
  {author} {\bibfnamefont {D.}~\bibnamefont {Sank}}, \bibinfo {author}
  {\bibfnamefont {H.}~\bibnamefont {Wang}}, \bibinfo {author} {\bibfnamefont
  {J.}~\bibnamefont {Wenner}}, \bibinfo {author} {\bibfnamefont {A.~N.}\
  \bibnamefont {Cleland}}, \bibinfo {author} {\bibfnamefont {M.~R.}\
  \bibnamefont {Geller}}, \ and\ \bibinfo {author} {\bibfnamefont {J.~M.}\
  \bibnamefont {Martinis}},\ }\href@noop {} {\bibfield  {journal} {\bibinfo
  {journal} {Science}\ }\textbf {\bibinfo {volume} {325}},\ \bibinfo {pages}
  {722} (\bibinfo {year} {2009})}\BibitemShut {NoStop}%
\bibitem [{\citenamefont {Lanyon}\ \emph {et~al.}(2009)\citenamefont {Lanyon},
  \citenamefont {Barbieri}, \citenamefont {Almeida}, \citenamefont {Jennewein},
  \citenamefont {Ralph}, \citenamefont {Resch}, \citenamefont {Pryde},
  \citenamefont {O'Brien}, \citenamefont {Gilchrist},\ and\ \citenamefont
  {White}}]{Lanyon:2009aa}%
  \BibitemOpen
  \bibfield  {author} {\bibinfo {author} {\bibfnamefont {B.~P.}\ \bibnamefont
  {Lanyon}}, \bibinfo {author} {\bibfnamefont {M.}~\bibnamefont {Barbieri}},
  \bibinfo {author} {\bibfnamefont {M.~P.}\ \bibnamefont {Almeida}}, \bibinfo
  {author} {\bibfnamefont {T.}~\bibnamefont {Jennewein}}, \bibinfo {author}
  {\bibfnamefont {T.~C.}\ \bibnamefont {Ralph}}, \bibinfo {author}
  {\bibfnamefont {K.~J.}\ \bibnamefont {Resch}}, \bibinfo {author}
  {\bibfnamefont {G.~J.}\ \bibnamefont {Pryde}}, \bibinfo {author}
  {\bibfnamefont {J.~L.}\ \bibnamefont {O'Brien}}, \bibinfo {author}
  {\bibfnamefont {A.}~\bibnamefont {Gilchrist}}, \ and\ \bibinfo {author}
  {\bibfnamefont {A.~G.}\ \bibnamefont {White}},\ }\href@noop {} {\bibfield
  {journal} {\bibinfo  {journal} {Nature Physics}\ }\textbf {\bibinfo {volume}
  {5}},\ \bibinfo {pages} {134} (\bibinfo {year} {2009})}\BibitemShut {NoStop}%
\bibitem [{\citenamefont {Bechmann-Pasquinucci}\ and\ \citenamefont
  {Peres}(2000)}]{Helle:2000}%
  \BibitemOpen
  \bibfield  {author} {\bibinfo {author} {\bibfnamefont {H.}~\bibnamefont
  {Bechmann-Pasquinucci}}\ and\ \bibinfo {author} {\bibfnamefont
  {A.}~\bibnamefont {Peres}},\ }\href@noop {} {\bibfield  {journal} {\bibinfo
  {journal} {Physical Review Letters}\ }\textbf {\bibinfo {volume} {85}},\
  \bibinfo {pages} {3313} (\bibinfo {year} {2000})}\BibitemShut {NoStop}%
\bibitem [{\citenamefont {Cerf}\ \emph {et~al.}(2002)\citenamefont {Cerf},
  \citenamefont {Bourennane}, \citenamefont {Karlsson},\ and\ \citenamefont
  {Gisin}}]{Cerf:2002}%
  \BibitemOpen
  \bibfield  {author} {\bibinfo {author} {\bibfnamefont {N.~J.}\ \bibnamefont
  {Cerf}}, \bibinfo {author} {\bibfnamefont {M.}~\bibnamefont {Bourennane}},
  \bibinfo {author} {\bibfnamefont {A.}~\bibnamefont {Karlsson}}, \ and\
  \bibinfo {author} {\bibfnamefont {N.}~\bibnamefont {Gisin}},\ }\href@noop {}
  {\bibfield  {journal} {\bibinfo  {journal} {Physical Review Letters}\
  }\textbf {\bibinfo {volume} {88}},\ \bibinfo {pages} {127902} (\bibinfo
  {year} {2002})}\BibitemShut {NoStop}%
\bibitem [{\citenamefont {Bianchetti}\ \emph {et~al.}(2010)\citenamefont
  {Bianchetti}, \citenamefont {Filipp}, \citenamefont {Baur}, \citenamefont
  {Fink}, \citenamefont {Lang}, \citenamefont {Steffen}, \citenamefont
  {Boissonneault}, \citenamefont {Blais},\ and\ \citenamefont
  {Wallraff}}]{Bianchetti:2010}%
  \BibitemOpen
  \bibfield  {author} {\bibinfo {author} {\bibfnamefont {R.}~\bibnamefont
  {Bianchetti}}, \bibinfo {author} {\bibfnamefont {S.}~\bibnamefont {Filipp}},
  \bibinfo {author} {\bibfnamefont {M.}~\bibnamefont {Baur}}, \bibinfo {author}
  {\bibfnamefont {J.~M.}\ \bibnamefont {Fink}}, \bibinfo {author}
  {\bibfnamefont {C.}~\bibnamefont {Lang}}, \bibinfo {author} {\bibfnamefont
  {L.}~\bibnamefont {Steffen}}, \bibinfo {author} {\bibfnamefont
  {M.}~\bibnamefont {Boissonneault}}, \bibinfo {author} {\bibfnamefont
  {A.}~\bibnamefont {Blais}}, \ and\ \bibinfo {author} {\bibfnamefont
  {A.}~\bibnamefont {Wallraff}},\ }\href@noop {} {\bibfield  {journal}
  {\bibinfo  {journal} {Physical Review Letters}\ }\textbf {\bibinfo {volume}
  {105}},\ \bibinfo {pages} {223601} (\bibinfo {year} {2010})}\BibitemShut
  {NoStop}%
\bibitem [{\citenamefont {Peterer}\ \emph {et~al.}(2015)\citenamefont
  {Peterer}, \citenamefont {Bader}, \citenamefont {Jin}, \citenamefont {Yan},
  \citenamefont {Kamal}, \citenamefont {Gudmundsen}, \citenamefont {Leek},
  \citenamefont {Orlando}, \citenamefont {Oliver},\ and\ \citenamefont
  {Gustavsson}}]{Peterer:2015}%
  \BibitemOpen
  \bibfield  {author} {\bibinfo {author} {\bibfnamefont {M.~J.}\ \bibnamefont
  {Peterer}}, \bibinfo {author} {\bibfnamefont {S.~J.}\ \bibnamefont {Bader}},
  \bibinfo {author} {\bibfnamefont {X.}~\bibnamefont {Jin}}, \bibinfo {author}
  {\bibfnamefont {F.}~\bibnamefont {Yan}}, \bibinfo {author} {\bibfnamefont
  {A.}~\bibnamefont {Kamal}}, \bibinfo {author} {\bibfnamefont {T.~J.}\
  \bibnamefont {Gudmundsen}}, \bibinfo {author} {\bibfnamefont {P.~J.}\
  \bibnamefont {Leek}}, \bibinfo {author} {\bibfnamefont {T.~P.}\ \bibnamefont
  {Orlando}}, \bibinfo {author} {\bibfnamefont {W.~D.}\ \bibnamefont {Oliver}},
  \ and\ \bibinfo {author} {\bibfnamefont {S.}~\bibnamefont {Gustavsson}},\
  }\href@noop {} {\bibfield  {journal} {\bibinfo  {journal} {Physical Review
  Letters}\ }\textbf {\bibinfo {volume} {114}},\ \bibinfo {pages} {010501}
  (\bibinfo {year} {2015})}\BibitemShut {NoStop}%
\bibitem [{\citenamefont {Rosenblum}\ \emph
  {et~al.}(2018{\natexlab{b}})\citenamefont {Rosenblum}, \citenamefont
  {Reinhold}, \citenamefont {Mirrahimi}, \citenamefont {Jiang}, \citenamefont
  {Frunzio},\ and\ \citenamefont {Schoelkopf}}]{Rosenblum266}%
  \BibitemOpen
  \bibfield  {author} {\bibinfo {author} {\bibfnamefont {S.}~\bibnamefont
  {Rosenblum}}, \bibinfo {author} {\bibfnamefont {P.}~\bibnamefont {Reinhold}},
  \bibinfo {author} {\bibfnamefont {M.}~\bibnamefont {Mirrahimi}}, \bibinfo
  {author} {\bibfnamefont {L.}~\bibnamefont {Jiang}}, \bibinfo {author}
  {\bibfnamefont {L.}~\bibnamefont {Frunzio}}, \ and\ \bibinfo {author}
  {\bibfnamefont {R.~J.}\ \bibnamefont {Schoelkopf}},\ }\href@noop {}
  {\bibfield  {journal} {\bibinfo  {journal} {Science}\ }\textbf {\bibinfo
  {volume} {361}},\ \bibinfo {pages} {266} (\bibinfo {year}
  {2018}{\natexlab{b}})}\BibitemShut {NoStop}%
\bibitem [{\citenamefont {Blok}\ \emph {et~al.}()\citenamefont {Blok},
  \citenamefont {Ramasesh}, \citenamefont {Schuster}, \citenamefont {O'Brien},
  \citenamefont {Kreikebaum}, \citenamefont {Dahlen}, \citenamefont {Morvan},
  \citenamefont {Yoshida}, \citenamefont {Yao},\ and\ \citenamefont
  {Siddiqi}}]{Blok:2020}%
  \BibitemOpen
  \bibfield  {author} {\bibinfo {author} {\bibfnamefont {M.~S.}\ \bibnamefont
  {Blok}}, \bibinfo {author} {\bibfnamefont {V.~V.}\ \bibnamefont {Ramasesh}},
  \bibinfo {author} {\bibfnamefont {T.}~\bibnamefont {Schuster}}, \bibinfo
  {author} {\bibfnamefont {K.}~\bibnamefont {O'Brien}}, \bibinfo {author}
  {\bibfnamefont {J.~M.}\ \bibnamefont {Kreikebaum}}, \bibinfo {author}
  {\bibfnamefont {D.}~\bibnamefont {Dahlen}}, \bibinfo {author} {\bibfnamefont
  {A.}~\bibnamefont {Morvan}}, \bibinfo {author} {\bibfnamefont
  {B.}~\bibnamefont {Yoshida}}, \bibinfo {author} {\bibfnamefont {N.~Y.}\
  \bibnamefont {Yao}}, \ and\ \bibinfo {author} {\bibfnamefont
  {I.}~\bibnamefont {Siddiqi}},\ }\href@noop {} {\bibinfo  {journal}
  {arXiv:2003.03307v1}\ }\BibitemShut {NoStop}%
\bibitem [{\citenamefont {Paik}\ \emph {et~al.}(2011)\citenamefont {Paik},
  \citenamefont {Schuster}, \citenamefont {Bishop}, \citenamefont {Kirchmair},
  \citenamefont {Catelani}, \citenamefont {Sears}, \citenamefont {Johnson},
  \citenamefont {Reagor}, \citenamefont {Frunzio}, \citenamefont {Glazman},
  \citenamefont {Girvin}, \citenamefont {Devoret},\ and\ \citenamefont
  {Schoelkopf}}]{Paik:2011}%
  \BibitemOpen
\bibfield  {journal} {  }\bibfield  {author} {\bibinfo {author} {\bibfnamefont
  {H.}~\bibnamefont {Paik}}, \bibinfo {author} {\bibfnamefont {D.~I.}\
  \bibnamefont {Schuster}}, \bibinfo {author} {\bibfnamefont {L.~S.}\
  \bibnamefont {Bishop}}, \bibinfo {author} {\bibfnamefont {G.}~\bibnamefont
  {Kirchmair}}, \bibinfo {author} {\bibfnamefont {G.}~\bibnamefont {Catelani}},
  \bibinfo {author} {\bibfnamefont {A.~P.}\ \bibnamefont {Sears}}, \bibinfo
  {author} {\bibfnamefont {B.~R.}\ \bibnamefont {Johnson}}, \bibinfo {author}
  {\bibfnamefont {M.~J.}\ \bibnamefont {Reagor}}, \bibinfo {author}
  {\bibfnamefont {L.}~\bibnamefont {Frunzio}}, \bibinfo {author} {\bibfnamefont
  {L.~I.}\ \bibnamefont {Glazman}}, \bibinfo {author} {\bibfnamefont {S.~M.}\
  \bibnamefont {Girvin}}, \bibinfo {author} {\bibfnamefont {M.~H.}\
  \bibnamefont {Devoret}}, \ and\ \bibinfo {author} {\bibfnamefont {R.~J.}\
  \bibnamefont {Schoelkopf}},\ }\href@noop {} {\bibfield  {journal} {\bibinfo
  {journal} {Physical Review Letters}\ }\textbf {\bibinfo {volume} {107}},\
  \bibinfo {pages} {240501} (\bibinfo {year} {2011})}\BibitemShut {NoStop}%
\bibitem [{\citenamefont {Lloyd}\ and\ \citenamefont
  {Montangero}(2014)}]{Lloyd:2014}%
  \BibitemOpen
  \bibfield  {author} {\bibinfo {author} {\bibfnamefont {S.}~\bibnamefont
  {Lloyd}}\ and\ \bibinfo {author} {\bibfnamefont {S.}~\bibnamefont
  {Montangero}},\ }\href@noop {} {\bibfield  {journal} {\bibinfo  {journal}
  {Physical Review Letters}\ }\textbf {\bibinfo {volume} {113}},\ \bibinfo
  {pages} {010502} (\bibinfo {year} {2014})}\BibitemShut {NoStop}%
\bibitem [{\citenamefont {Blais}\ \emph {et~al.}(2004)\citenamefont {Blais},
  \citenamefont {Huang}, \citenamefont {Wallraff}, \citenamefont {Girvin},\
  and\ \citenamefont {Schoelkopf}}]{Blais:2004}%
  \BibitemOpen
  \bibfield  {author} {\bibinfo {author} {\bibfnamefont {A.}~\bibnamefont
  {Blais}}, \bibinfo {author} {\bibfnamefont {R.-S.}\ \bibnamefont {Huang}},
  \bibinfo {author} {\bibfnamefont {A.}~\bibnamefont {Wallraff}}, \bibinfo
  {author} {\bibfnamefont {S.~M.}\ \bibnamefont {Girvin}}, \ and\ \bibinfo
  {author} {\bibfnamefont {R.~J.}\ \bibnamefont {Schoelkopf}},\ }\href@noop {}
  {\bibfield  {journal} {\bibinfo  {journal} {Physical Review A}\ }\textbf
  {\bibinfo {volume} {69}},\ \bibinfo {pages} {062320} (\bibinfo {year}
  {2004})}\BibitemShut {NoStop}%
\bibitem [{\citenamefont {Gerry}\ and\ \citenamefont {Knight}(2004)}]{Gerry}%
  \BibitemOpen
  \bibfield  {author} {\bibinfo {author} {\bibfnamefont {C.}~\bibnamefont
  {Gerry}}\ and\ \bibinfo {author} {\bibfnamefont {P.}~\bibnamefont {Knight}},\
  }\href@noop {} {\emph {\bibinfo {title} {Introductory Quantum Optics}}}\
  (\bibinfo  {publisher} {Cambridge University Press},\ \bibinfo {year}
  {2004})\BibitemShut {NoStop}%
\bibitem [{Note1()}]{Note1}%
  \BibitemOpen
  \bibinfo {note} {See the supplemental material at [PRL LINK] which contains
  Refs.~\protect \rev@citealp {Blais:2004,scikit-learn, Reagor:2018,Gerry} for
  details of the model and calculation.}\BibitemShut {Stop}%
\bibitem [{\citenamefont {R.~Johansson}\ \emph {et~al.}(2013)\citenamefont
  {R.~Johansson}, \citenamefont {Nation},\ and\ \citenamefont
  {Nori}}]{Johansson:2013}%
  \BibitemOpen
  \bibfield  {author} {\bibinfo {author} {\bibfnamefont {J.}~\bibnamefont
  {R.~Johansson}}, \bibinfo {author} {\bibfnamefont {P.~D.}\ \bibnamefont
  {Nation}}, \ and\ \bibinfo {author} {\bibfnamefont {F.}~\bibnamefont
  {Nori}},\ }\href@noop {} {\bibfield  {journal} {\bibinfo  {journal} {Computer
  Physics Communications}\ }\textbf {\bibinfo {volume} {184}},\ \bibinfo
  {pages} {1234} (\bibinfo {year} {2013})}\BibitemShut {NoStop}%
\bibitem [{\citenamefont {Leung}\ \emph {et~al.}(2017)\citenamefont {Leung},
  \citenamefont {Abdelhafez}, \citenamefont {Koch},\ and\ \citenamefont
  {Schuster}}]{Leung:2017}%
  \BibitemOpen
  \bibfield  {author} {\bibinfo {author} {\bibfnamefont {N.}~\bibnamefont
  {Leung}}, \bibinfo {author} {\bibfnamefont {M.}~\bibnamefont {Abdelhafez}},
  \bibinfo {author} {\bibfnamefont {J.}~\bibnamefont {Koch}}, \ and\ \bibinfo
  {author} {\bibfnamefont {D.}~\bibnamefont {Schuster}},\ }\href@noop {}
  {\bibfield  {journal} {\bibinfo  {journal} {Physical Review A}\ }\textbf
  {\bibinfo {volume} {95}},\ \bibinfo {pages} {042318} (\bibinfo {year}
  {2017})}\BibitemShut {NoStop}%
\bibitem [{\citenamefont {de~Fouquieres}\ \emph {et~al.}(2011)\citenamefont
  {de~Fouquieres}, \citenamefont {Schirmer}, \citenamefont {Glaser},\ and\
  \citenamefont {Kuprov}}]{Fouquieres:2011}%
  \BibitemOpen
  \bibfield  {author} {\bibinfo {author} {\bibfnamefont {P.}~\bibnamefont
  {de~Fouquieres}}, \bibinfo {author} {\bibfnamefont {S.~G.}\ \bibnamefont
  {Schirmer}}, \bibinfo {author} {\bibfnamefont {S.~J.}\ \bibnamefont
  {Glaser}}, \ and\ \bibinfo {author} {\bibfnamefont {L.}~\bibnamefont
  {Kuprov}},\ }\href@noop {} {\bibfield  {journal} {\bibinfo  {journal}
  {Journal of Magnetic Resonance}\ }\textbf {\bibinfo {volume} {212}},\
  \bibinfo {pages} {412} (\bibinfo {year} {2011})}\BibitemShut {NoStop}%
\bibitem [{\citenamefont {Petersson}\ \emph {et~al.}()\citenamefont
  {Petersson}, \citenamefont {Fortino}, \citenamefont {Copeland}, \citenamefont
  {Rydin},\ and\ \citenamefont {L~DuBois}}]{anders:2020}%
  \BibitemOpen
  \bibfield  {author} {\bibinfo {author} {\bibfnamefont {N.~A.}\ \bibnamefont
  {Petersson}}, \bibinfo {author} {\bibfnamefont {G.~M.}\ \bibnamefont
  {Fortino}}, \bibinfo {author} {\bibfnamefont {A.~E.}\ \bibnamefont
  {Copeland}}, \bibinfo {author} {\bibfnamefont {Y.~L.}\ \bibnamefont {Rydin}},
  \ and\ \bibinfo {author} {\bibfnamefont {J.}~\bibnamefont {L~DuBois}},\
  }\href@noop {} {\bibinfo  {journal} {arXiv:2001.01013}\ }\BibitemShut
  {NoStop}%
\bibitem [{\citenamefont {Goran}(1976)}]{Lindblad}%
  \BibitemOpen
\bibfield  {journal} {  }\bibfield  {author} {\bibinfo {author} {\bibfnamefont
  {L.}~\bibnamefont {Goran}},\ }\href@noop {} {\bibfield  {journal} {\bibinfo
  {journal} {Communications in Mathematical Physics}\ }\textbf {\bibinfo
  {volume} {48}},\ \bibinfo {pages} {119} (\bibinfo {year} {1976})}\BibitemShut
  {NoStop}%
\bibitem [{\citenamefont {Goupillaud}\ \emph {et~al.}(1984)\citenamefont
  {Goupillaud}, \citenamefont {Grossmann},\ and\ \citenamefont
  {Morlet}}]{Morlet}%
  \BibitemOpen
  \bibfield  {author} {\bibinfo {author} {\bibfnamefont {P.}~\bibnamefont
  {Goupillaud}}, \bibinfo {author} {\bibfnamefont {A.}~\bibnamefont
  {Grossmann}}, \ and\ \bibinfo {author} {\bibfnamefont {J.}~\bibnamefont
  {Morlet}},\ }\href@noop {} {\bibfield  {journal} {\bibinfo  {journal}
  {Geoexploration}\ }\textbf {\bibinfo {volume} {23}},\ \bibinfo {pages} {85}
  (\bibinfo {year} {1984})}\BibitemShut {NoStop}%
\bibitem [{\citenamefont {Kumar}\ \emph {et~al.}(2016)\citenamefont {Kumar},
  \citenamefont {Veps{\"a}l{\"a}inen}, \citenamefont {Danilin},\ and\
  \citenamefont {Paraoanu}}]{Kumar:2016aa}%
  \BibitemOpen
  \bibfield  {author} {\bibinfo {author} {\bibfnamefont {K.~S.}\ \bibnamefont
  {Kumar}}, \bibinfo {author} {\bibfnamefont {A.}~\bibnamefont
  {Veps{\"a}l{\"a}inen}}, \bibinfo {author} {\bibfnamefont {S.}~\bibnamefont
  {Danilin}}, \ and\ \bibinfo {author} {\bibfnamefont {G.~S.}\ \bibnamefont
  {Paraoanu}},\ }\href {\doibase 10.1038/ncomms10628} {\bibfield  {journal}
  {\bibinfo  {journal} {Nature Communications}\ }\textbf {\bibinfo {volume}
  {7}},\ \bibinfo {pages} {10628} (\bibinfo {year} {2016})}\BibitemShut
  {NoStop}%
\bibitem [{\citenamefont {Xu}\ \emph {et~al.}(2016)\citenamefont {Xu},
  \citenamefont {Song}, \citenamefont {Liu}, \citenamefont {Xue}, \citenamefont
  {Su}, \citenamefont {Deng}, \citenamefont {Tian}, \citenamefont {Zheng},
  \citenamefont {Han}, \citenamefont {Zhong}, \citenamefont {Wang},
  \citenamefont {Liu},\ and\ \citenamefont {Zhao}}]{Xu:2016aa}%
  \BibitemOpen
  \bibfield  {author} {\bibinfo {author} {\bibfnamefont {H.~K.}\ \bibnamefont
  {Xu}}, \bibinfo {author} {\bibfnamefont {C.}~\bibnamefont {Song}}, \bibinfo
  {author} {\bibfnamefont {W.~Y.}\ \bibnamefont {Liu}}, \bibinfo {author}
  {\bibfnamefont {G.~M.}\ \bibnamefont {Xue}}, \bibinfo {author} {\bibfnamefont
  {F.~F.}\ \bibnamefont {Su}}, \bibinfo {author} {\bibfnamefont
  {H.}~\bibnamefont {Deng}}, \bibinfo {author} {\bibfnamefont {Y.}~\bibnamefont
  {Tian}}, \bibinfo {author} {\bibfnamefont {D.~N.}\ \bibnamefont {Zheng}},
  \bibinfo {author} {\bibfnamefont {S.}~\bibnamefont {Han}}, \bibinfo {author}
  {\bibfnamefont {Y.~P.}\ \bibnamefont {Zhong}}, \bibinfo {author}
  {\bibfnamefont {H.}~\bibnamefont {Wang}}, \bibinfo {author} {\bibfnamefont
  {Y.-x.}\ \bibnamefont {Liu}}, \ and\ \bibinfo {author} {\bibfnamefont
  {S.~P.}\ \bibnamefont {Zhao}},\ }\href {\doibase 10.1038/ncomms11018}
  {\bibfield  {journal} {\bibinfo  {journal} {Nature Communications}\ }\textbf
  {\bibinfo {volume} {7}},\ \bibinfo {pages} {11018} (\bibinfo {year}
  {2016})}\BibitemShut {NoStop}%
\bibitem [{\citenamefont {L.~Chuang}\ and\ \citenamefont
  {A.~Nielsen}(1997)}]{QPT}%
  \BibitemOpen
  \bibfield  {author} {\bibinfo {author} {\bibfnamefont {I.}~\bibnamefont
  {L.~Chuang}}\ and\ \bibinfo {author} {\bibfnamefont {M.}~\bibnamefont
  {A.~Nielsen}},\ }\href@noop {} {\bibfield  {journal} {\bibinfo  {journal}
  {Journal of Modern Optics}\ }\textbf {\bibinfo {volume} {44}} (\bibinfo
  {year} {1997})}\BibitemShut {NoStop}%
\bibitem [{\citenamefont {Chow}\ \emph {et~al.}(2009)\citenamefont {Chow},
  \citenamefont {Gambetta}, \citenamefont {Tornberg}, \citenamefont {Koch},
  \citenamefont {Bishop}, \citenamefont {Houck}, \citenamefont {Johnson},
  \citenamefont {Frunzio}, \citenamefont {Girvin},\ and\ \citenamefont
  {Schoelkopf}}]{Chow:2009}%
  \BibitemOpen
  \bibfield  {author} {\bibinfo {author} {\bibfnamefont {J.~M.}\ \bibnamefont
  {Chow}}, \bibinfo {author} {\bibfnamefont {J.~M.}\ \bibnamefont {Gambetta}},
  \bibinfo {author} {\bibfnamefont {L.}~\bibnamefont {Tornberg}}, \bibinfo
  {author} {\bibfnamefont {J.}~\bibnamefont {Koch}}, \bibinfo {author}
  {\bibfnamefont {L.~S.}\ \bibnamefont {Bishop}}, \bibinfo {author}
  {\bibfnamefont {A.~A.}\ \bibnamefont {Houck}}, \bibinfo {author}
  {\bibfnamefont {B.~R.}\ \bibnamefont {Johnson}}, \bibinfo {author}
  {\bibfnamefont {L.}~\bibnamefont {Frunzio}}, \bibinfo {author} {\bibfnamefont
  {S.~M.}\ \bibnamefont {Girvin}}, \ and\ \bibinfo {author} {\bibfnamefont
  {R.~J.}\ \bibnamefont {Schoelkopf}},\ }\href@noop {} {\bibfield  {journal}
  {\bibinfo  {journal} {Physical Review Letters}\ }\textbf {\bibinfo {volume}
  {102}},\ \bibinfo {pages} {090502} (\bibinfo {year} {2009})}\BibitemShut
  {NoStop}%
\bibitem [{\citenamefont {Chow}\ \emph {et~al.}(2010)\citenamefont {Chow},
  \citenamefont {DiCarlo}, \citenamefont {Gambetta}, \citenamefont {Motzoi},
  \citenamefont {Frunzio}, \citenamefont {Girvin},\ and\ \citenamefont
  {Schoelkopf}}]{PhysRevA.82.040305}%
  \BibitemOpen
  \bibfield  {author} {\bibinfo {author} {\bibfnamefont {J.~M.}\ \bibnamefont
  {Chow}}, \bibinfo {author} {\bibfnamefont {L.}~\bibnamefont {DiCarlo}},
  \bibinfo {author} {\bibfnamefont {J.~M.}\ \bibnamefont {Gambetta}}, \bibinfo
  {author} {\bibfnamefont {F.}~\bibnamefont {Motzoi}}, \bibinfo {author}
  {\bibfnamefont {L.}~\bibnamefont {Frunzio}}, \bibinfo {author} {\bibfnamefont
  {S.~M.}\ \bibnamefont {Girvin}}, \ and\ \bibinfo {author} {\bibfnamefont
  {R.~J.}\ \bibnamefont {Schoelkopf}},\ }\href@noop {} {\bibfield  {journal}
  {\bibinfo  {journal} {Physical Review A}\ }\textbf {\bibinfo {volume} {82}},\
  \bibinfo {pages} {040305(R)} (\bibinfo {year} {2010})}\BibitemShut {NoStop}%
\end{thebibliography}%

\end{document}